\documentstyle[twocolumn]{mn}

\def\beq{\begin{equation}}
\def\eeq{\end{equation}}
\def\bey{\begin{eqnarray}}
\def\eey{\end{eqnarray}}
\def\kms{\,{\rm km}\,{\rm s}^{-1}}
\def\msunpc2{{\rm \,M}_\odot\,{\rm pc}^{-2}}
\def\mas{{\rm \,mas}}
\def\AU{{\rm \,au}}

\def\thetaE{\theta_{\rm E}}
\def\RE{R_{\rm E}}
\def\rE{{{\tilde r}_{\rm E}}}

\def\Tlife{T_{\rm life}}
\def\tE{t_{\rm E}}
\def\tae{t_{\rm ae}}
\def\up{u_{\rm p}}
\def\ua{u_{\rm a}}
\def\Rd{R_{\rm d}}
\def\zh{z_{\rm h}}
\def\ms{m_{\rm G}}
\def\Ms{M_{G}}
\def\Ds{D_{\rm s}}
\def\Dl{D_{\rm l}}

\def\msun{M_\odot}
\def\Msun{M_\odot}
\def\ds{\displaystyle}

\def\vmean{{v_{\rm m}}}
\def\that{{\hat{t}}}
\def\btheta{\mbox{\boldmath $\theta$}}
\def\bmu{\mbox{\boldmath $\mu$}}
\def\muas{\;\mu{\rm as}}
\def\tmuas{\mu{\rm as}}
\def\taua{\tau_{\rm a}}
\def\taup{\tau_{\rm p}}
\def\sigmaa{\sigma_{\rm a}}

\def\sigmap{\sigma_{\rm p}}
\def\pc{{\rm pc}}
\def\spose#1{\hbox to 0pt{#1\hss}}
\def\lta{\mathrel{\spose{\lower 3pt\hbox{$\sim$}}
    \raise 2.0pt\hbox{$<$}}}
\def\gta{\mathrel{\spose{\lower 3pt\hbox{$\sim$}}
    \raise 2.0pt\hbox{$>$}}}

\input epsf

\title[Astrometric Microlensing with the GAIA satellite]
      {Astrometric Microlensing with the GAIA satellite}
\author[V.A. Belokurov \& N.W. Evans]
        {V.A. Belokurov \& N.W. Evans \\
        Theoretical Physics, 1 Keble Rd, Oxford, OX1 3NP}

\begin{document}
\maketitle
\label{firstpage}

\begin{abstract}
GAIA is the ``super-Hipparcos'' survey satellite selected as a
Cornerstone 6 mission by the European Space Agency. GAIA can measure
microlensing by the brightening of source stars.  For the broad G band
photometer, the all-sky source-averaged photometric optical depth is
$\sim 10^{-7}$.  There are $\sim 1300$ photometric microlensing events
for which GAIA will measure at least one datapoint on the amplified
lightcurve.  GAIA can also measure microlensing by the small
excursions of the light centroid that occur during events.  The
all-sky source-averaged astrometric microlensing optical depth is
$\sim 2.5 \times 10^{-5}$. Some $\sim 25000$ sources will have a
significant variation of the centroid shift, together with a closest
approach, during the lifetime of the mission. This is not the actual
number of events that can be extracted from the GAIA dataset, as the
false detection rate has not been assessed.

A covariance analysis is used to study the propagation of errors and
the estimation of parameters from realistic sampling of the GAIA
datastream of transits in the along-scan direction during microlensing
events.  The mass of the lens can be calculated to good accuracy if
the lens is nearby so that angular Einstein radius $\thetaE$ is large;
if the Einstein radius projected onto the observer plane $\rE$ is
about an astronomical unit; if the duration of the astrometric event
is long ($\gta 1$ year) or if the source star is bright.  Monte Carlo
simulations are used to study the $\sim 2500$ events for which the
mass can be recovered with an error of $< 50 \%$. These high quality
events are dominated by disk lenses within a few tens of parsecs and
source stars within a few hundred parsecs. We show that the local mass
function can be recovered from the high quality sample to good
accuracy. GAIA is the first instrument with the capabilities of
measuring the mass locally in very faint objects like black holes and
very cool white and brown dwarfs.

For only $\sim 5\%$ of all astrometric events will GAIA record even
one photometric datapoint. There is a need for a dedicated telescope
that densely samples the Galactic Centre and spiral arms, as this can
improve the accuracy of parameter estimation by a factor of $\sim
10$. The total number of sources that have an astrometric microlensing
variation exceeding the mission target accuracy is $\sim 10^5$.  The
positional measurement of one source in every twenty thousand is affected
by microlensing noise at any instant. We show that microlensing is
negligible as an unbiased random error source for GAIA.
\end{abstract}


\section{Introduction}

GAIA is the European Space Agency (ESA) satellite now selected as a
Cornerstone 6 mission as part of the Science
Program~\footnote{http://astro.estec.esa.nl/gaia}. It is a survey
satellite that provides multi-colour, multi-epoch photometry,
astrometry and spectroscopy on all objects brighter then $V\approx 20$
(e.g., ESA 2000; Perryman et al. 2001).  The dataset is huge with
information of unprecedented precision on over a billion objects in
our Galaxy alone.

GAIA is the successor to the pioneering {\it Hipparcos} satellite,
which flew from 1989 to 1993. The {\it Hipparcos} program was both
simpler and smaller: it measured only $10^5$ rather than $10^9$
objects and it was provided with a target list rather than being left
to determine its targets for itself, as GAIA will. The {\it Hipparcos}
final results were released only when the mission was complete,
whereas many of the science goals of GAIA, especially for bursting or
time-varying phenomena like microlensing, may require an early analysis
and release of preliminary data.

GAIA carries out continuous scanning of the sky. The satellite rotates
slowly on its spin axis, which itself precesses at a fixed angle to
the Sun of $55^\circ$.  As GAIA rotates, light enters the entrance
chambers, reflects off mirrors and falls on the focal planes of one of
three telescopes, two of which (ASTRO-1, ASTRO-2) measure the
positions of stars, one of which (SPECTRO) performs spectroscopy. GAIA
observes in three directions along a great circle
simultaneously. The astrometric parameters of stars are recovered from
the time series of the one-dimensional transits distributed over the
five year mission lifetime.

A small fraction of the objects monitored by GAIA will show evidence
of microlensing. GAIA can observe microlensing by measuring the
photometric amplification of a source star when a lens and a source
are aligned.  This is the approach followed by the large ground-based
microlensing surveys like MACHO, EROS, OGLE and POINT-AGAPE (see e.g.,
Alcock et al. 1997; Aubourg et al. 1995; Udalski et al. 1994,
Auri\`ere et al. 2001). GAIA is inefficient at discovering photometric
microlensing events, as the sampling of individual objects is
relatively sparse (there are a cluster of observations once every two
months on average).

\begin{figure}
\epsfxsize=8cm \centerline{\epsfbox{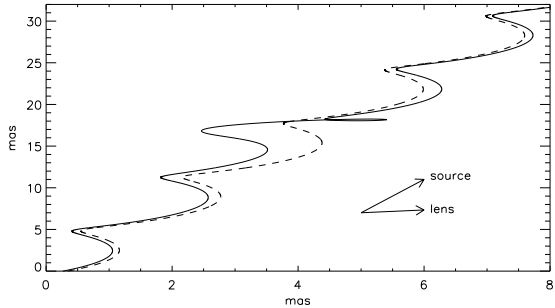}}
\caption{This shows the relative right ascension and declination of a
source with (solid line) and without (dashed line) a microlensing
event. The trajectory is shown over the GAIA mission lifetime of 5
years.  Note that the deviations caused by microlensing are present
well before and after the time of maximum of the event. (The lens is
at 150 pc, the source at 1.5 kpc, the transverse velocity is $70 \kms$
and the impact parameter $u$ is 1.5. The lens has mass $0.5\msun$.)}
\label{fig:pathofsource}
\end{figure}

However, there is a much more powerful strategy available to GAIA.
Although the two images of a microlensed source are unresolvable, GAIA
can measure the small deviation (of the order of a fraction of a
milliarcsec) of the centroid of the two images around the trajectory
of the source. Astrometric microlensing is the name given to this
excursion of the image centroid.  The cross section of a lens is
proportional to the area it sweeps out on the sky, and so to the
product of lens proper motion and angular Einstein radius. Each of
these varies as the inverse square root of lens distance, so the
signal is dominated by nearby lenses.  The detection of astrometric
microlensing events for pointed observations has been considered many
times before (e.g., Walker 1995; Miralda-Escud\'e 1996; Paczy\'nski
1996; Boden, Shao \& van Buren 1998). Once alerted by a ground-based
survey, the events are followed by narrow angle differential
astrometry. The detection of astrometric microlensing using
observations from a scanning satellite has been considered before by
H{\o}g, Novikov \& Polnarev (1995) in the context of the proposed
ROEMER mission (a forerunner of GAIA).

We will show that the all-sky source-averaged astrometric microlensing
optical depth is $\sim 2.5 \times 10^{-5}$, over an order of magnitude
greater than the photometric microlensing optical depth. There are two
main difficulties facing GAIA in exploiting this comparatively high
probability. First, the astrometric accuracy of a single measurement
by GAIA depends on the source magnitude and quickly degrades at
magnitudes fainter than $G \approx 15$.  Second, GAIA provides a
time-series of one-dimensional astrometry.  The observed quantity is
the CCD transit time for the coordinate along the scan. This is the
same way the Hipparcos satellite worked (see {\it The Hipparcos and
Tycho Catalogues}, ESA (1997), volume 3, section 16). From the
sequence of these one-dimensional measurements, the astrometric path
of the source, together with any additional deflection caused by
microlensing, must be recovered.

This paper assesses the microlensing signal that will be seen by GAIA.
Section 2 gives the formulae for the photometric and astrometric
microlensing optical depths. These are used to estimate the total
number of microlensing events that GAIA can measure. In Section 3, a
covariance analysis is used to demonstrate the effects of error
propagation on the recovery of the parameters of microlensing
events. For the purposes of GAIA, we show that the disk stars within a
few hundred parsecs of the Sun are the most important source and lens
populations.  Section 4 describes the extinction law and the Galaxy
model used to generate microlensing events for our Monte Carlo
simulations. The synthetic data are sampled with GAIA's scanning law
and realistic errors are applied to provide the one-dimensional
astrometric datastreams. Section 5 discusses the results of the
simulations, both for the entire sample of events and for the subset
of gold-plated events whose parameters can be recovered to good
accuracy. Finally, Section 6 examines the overall strategy for
identification of events, which need to be distinguished from other
forms of astrometric deviation such as binary companions.

\begin{figure}
\epsfxsize=8cm \centerline{\epsfbox{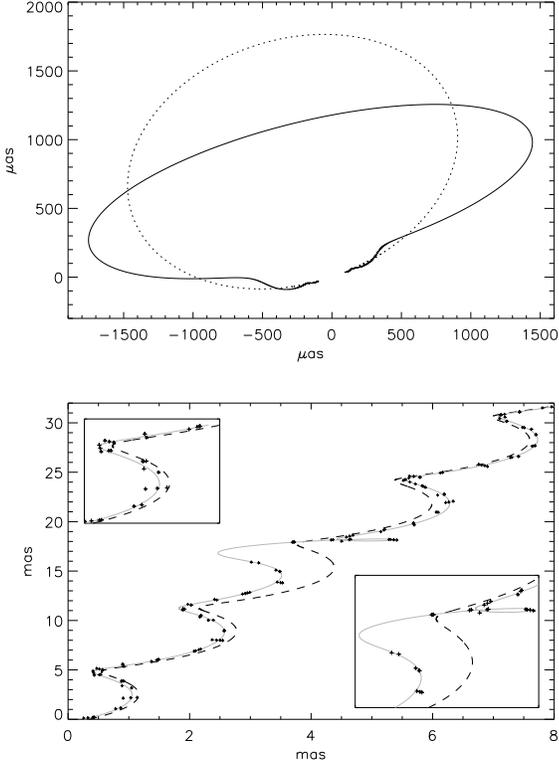}}
\caption{Upper panel: Astrometric shift of the microlensing event of
Figure 1, as seen by a barycentric (dotted line) and a terrestrial
observer (solid line).  Lower panel: Simulated data incorporating
typical sampling and astrometric errors for GAIA. Also shown for
comparison are the theoretical trajectories of the source with (grey
line) and without (dashed line) the event. The insets show the
deviations at the beginning and the midpoint of this high
signal-to-noise event. (The accuracy $\sigmaa$ of the astrometry is
$300\muas$, corresponding roughly to a $17$th magnitude star).}
\label{fig:examplesone}
\end{figure}
\begin{figure}
\epsfxsize=8cm \centerline{\epsfbox{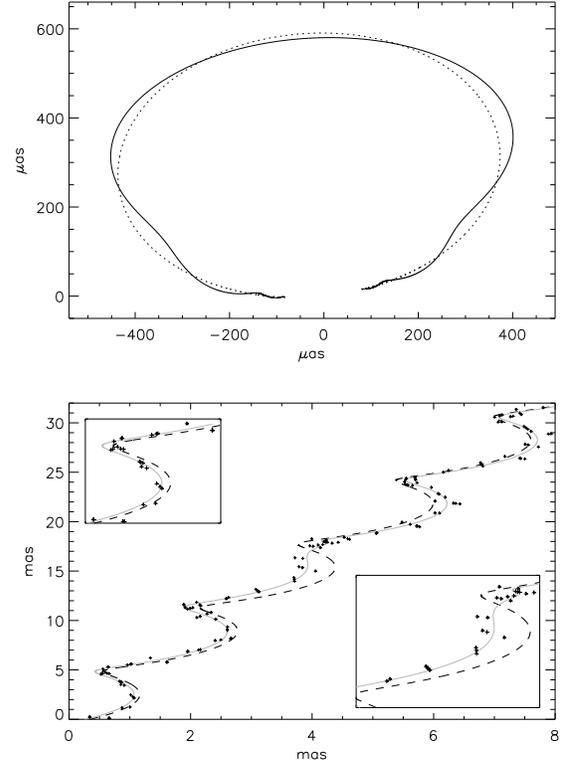}}
\caption{Same as Figure 2, but the lens distance is 1 kpc.  This shows
a microlensing event with a $5 \sqrt{2} \sigmaa$ variation of the
centroid shift. GAIA will be able to measure the astrometric deviation
but will not be able to extract any useful information on the
microlensing parameters.}
\label{fig:examplestwo}
\end{figure}
\begin{table*}
\begin{center}
\begin{tabular}{c|cccccccccccc}\hline
$G$ (in mag) & 10 & 11 & 12 & 13  & 14 & 15 & 16 & 17 & 18 & 19 & 20
\\
$\sigmap$ (in mmag) & 9 & 9 & 9 & 9 & 9 & 10 & 12 & 16 & 23 &
36 & 60 \\
$\sigmaa$ (in $\mu$as) & 30 & 30 & 30 & 40 & 60 & 90 & 150 & 230 & 390 &
700 & 1400 \\
\hline
\end{tabular}
\end{center}
\caption{This table lists the mean accuracy in photometry $\sigmap$
and in position $\sigmaa$ versus $G$ band magnitude for the GAIA
satellite. Note that $\sigmaa$ is the accuracy of a single astrometric
measurement, not the target accuracy at the end of the GAIA mission,
which is better by a factor of $\sim 10$. The values are approximate
sky averages, adapted from Tables 7.3 and 8.2 of ESA (2000).}
\label{table:astroacc}
\end{table*}
\begin{figure}
\epsfysize=15.cm \centerline{\epsfbox{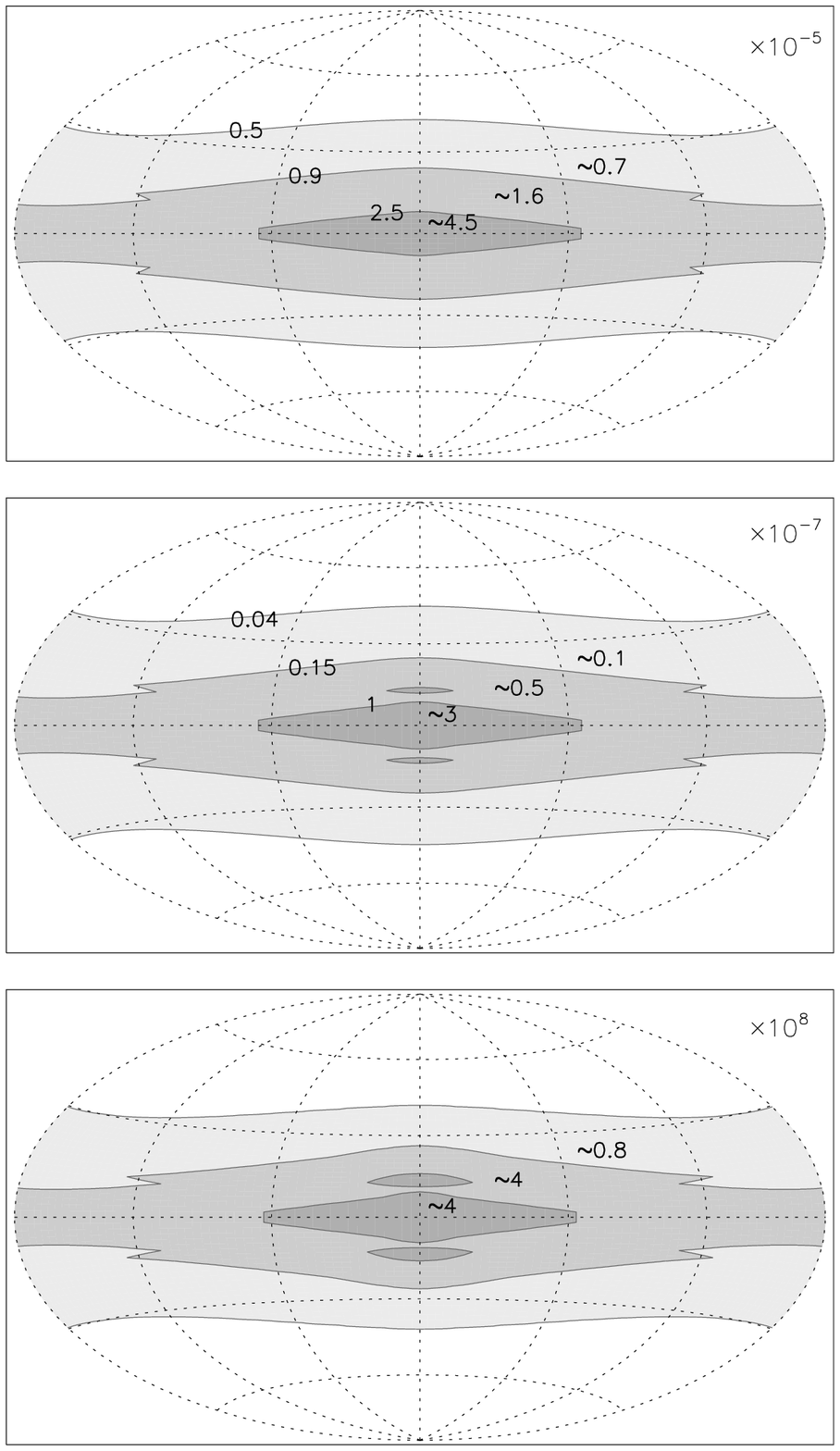}}
\caption{Upper panel: An all-sky map of the source-averaged
astrometric optical depth. Meridians of galactic latitude are shown at
$60^\circ$ intervals, parallels of longitude at $30^\circ$ intervals.
The contour levels are marked. The number within each contour refers
to the mean value of the optical depth inside the shaded region
enclosed between the contours. All numbers are in units of $10^{-5}$.
Middle panel: As the upper panel, but for the photometric microlensing
optical depth.  All numbers are in units of $10^{-7}$. Lower panel: An
all-sky map of the starcounts.  The number within each contour refers
to the total number of stars inside the shaded region enclosed between
the contours. To find the instantaneous number of events in any
region, we multiply the number of stars by the average optical depth.
(For details of the source and lens populations and extinction model,
see Section 4).}
\label{fig:maps}
\end{figure}

\section{Astrometric Microlensing}

\subsection{General Principles}

Let us consider the lensing of a luminous source by a dark point
lens. (Of course, the lens may be luminous, but we defer consideration
of this till Section 6). The angular position of the source
${\btheta}_{\rm s}$ can be written as
\beq
{\btheta}_{\rm s}(t) = {\btheta}_{{\rm s},0}
+ {\bmu}_{\rm s} t + {\bf P}_{\rm s},
\eeq
where ${\btheta}_{{\rm s},0}$ is the zero-point, ${\bmu}_{\rm s}$
is the source proper motion and ${\bf P}_{\rm s}$ is the source
parallax.  Similarly, the angular separation between source and lens
is given by
\beq
{\btheta}_{\rm sl}(t) = {\btheta}_{{\rm sl},0} 
+ {\bmu}_{\rm sl} t + {\bf P}_{\rm sl},
\eeq
where the proper motion ${\bmu}_{\rm sl}$ and parallax ${\bf P}_{\rm
sl}$ are of the source relative to the lens.

Microlensing induces an additional shift of the source centroid.
Letting ${\bf S}(t)$ be the image centroid, we have (e.g., Walker
1995; Dominik \& Sahu 2000)
\beq\label{centroid}
{\bf S}(t)= {\btheta}_{\rm s}(t) + {\btheta}(t),\qquad\qquad
{\btheta} = {{\btheta}_{\rm sl}(t) \over (\theta_{\rm sl}
(t)/\thetaE)^2 +2 }.
\eeq
Here, the angular Einstein radius $\thetaE$ is related to the lens
mass $M$ by
\beq {\thetaE \over \mas} = \left({M \over 0.12
M_\odot}\right)^{1/2} \left( {\pi_{\rm sl} \over \mas} \right)^{1/2}, 
\label{eq:thetae}
\eeq
where $\pi_{\rm sl} = |{\bf P}_{\rm sl}|$.
Figure~\ref{fig:pathofsource} shows the astrometric trajectory of the
same source with and without a microlensing event. In this case, the
lens is just 150 pc away from the observer, while the source is a disk
star 1.5 kpc away.  The microlensing event causes the additional
excursion superimposed on the parallactic and proper motion of the
source.

The characteristic lengthscale in microlensing is the Einstein radius
$\RE = \Dl \thetaE$. It is the linear size of the angular Einstein
radius in the lens plane. For applications in the Galaxy, we find that
\beq
\RE = 9 \AU \sqrt{ {M \over \Msun}} \sqrt{ \Dl \over 10 {\rm \;kpc}}
\sqrt{ 1 - {\Dl\over \Ds}},
\label{eq:einsteinr}
\eeq
and so $\RE$ is of the typical size of a few astronomical units
(aus). Here, $\Dl$ and $\Ds$ are the distances from the observer to
the lens and source respectively. The characteristic timescale in
photometric microlensing is the Einstein crossing time $\tE$, which is
the time taken for the source to cross the Einstein radius.

\subsection{The Centroid Shift}

The centroid shift points away from the dark lens as seen by the
source. The dimensionless function
\beq
{\bf u} = {{\btheta}_{\rm sl} \over \thetaE}
\eeq
is the angular separation in units of the angular Einstein radius.
For $u \rightarrow \infty$, the centroid shift falls off like
\beq
S = | {\bf S} |  \sim {\thetaE \over u}.
\eeq
This can be compared with the photometric magnification $A$, which
falls off like
\beq 
A \sim 1 + {2\over u^4}.
\eeq
These asymptotic results illustrate one of the important differences
between astrometric and photometric microlensing. The centroid shift
falls off much more slowly than the magnification, so that the
cross-section for astrometric events is much larger than for
photometric events (e.g., Paczy\'nski 1996; Miralda-Escud\'e 1996).

In the absence of the proper and parallactic motion, the absolute
value of the centroid shift is
\beq
\theta = {\sqrt{u_0^2 + \that^2} \over u_0^2 + \that^2 + 2} \thetaE.
\eeq
Here, $u_0$ is the value of the dimensionless distance at the time of
closest approach $t_0$ and
\beq
\that(t) = { t - t_0\over \tE} = {\mu_{\rm sl}  (t- t_0) \over \thetaE}. 
\eeq
The shift can be decomposed into components parallel and perpendicular
to the direction of motion of the lens relative to the source, namely
\begin{equation}
\theta_\parallel = {\that\over u_0^2 + \that^2 +2} \thetaE, \qquad\qquad
\theta_\perp = {u_0\over u_0^2 + \that^2 +2} \thetaE. 
\label{eq:walker}
\end{equation}
As pointed out by Walker (1995), the centroid shift as seen by a
barycentric observer is an ellipse. For $u \rightarrow 0$, the ellipse
becomes a straight line, while for $u \rightarrow \infty$, it becomes
a circle.

Figure~\ref{fig:examplesone} shows two further views of the same
microlensing event presented in Figure \ref{fig:pathofsource}. The
upper panel shows the right ascension and declination recorded by a
barycentric and a terrestrial observer (or equivalently a satellite at
the ${\rm L}_2$ Lagrange point, like GAIA). The proper and parallactic
motion of the source have been subtracted out. However, GAIA does not
provide the data in such a clean form, as it really measures a series
of one-dimensional transits of the two-dimensional astrometric
curve. The lower panel shows the event as seen by GAIA. The simulated
datapoints have been produced by generating random transit angles, and
sampling the astrometric curve according to GAIA's scanning law for
the ASTRO-1 and ASTRO-2 telescopes (using programs freely available
from L. Lindegren at Lund Observatory). The transits are strongly
clustered, as GAIA spins on its axis once every 3 hours and so may
scan the same patch of sky four or five times a day. The transit angle
is the same for all transits in such a cluster, but changes randomly
from cluster to cluster.  Gaussian astrometry errors with standard
deviation $\sigmaa = 300 \muas$ have been added to the simulated
datapoints.  The two insets show the astrometric deviations at the
beginning and at the maximum of the event, from which it is clear that
GAIA can detect that a microlensing event has
occurred. Figure~\ref{fig:examplestwo} shows the same microlensing
event, except that the lens distance has been increased.  This causes
the relative parallax $\pi_{\rm sl}$, and consequently the angular
Einstein radius $\thetaE$, to diminish. The upper panel of
Figure~\ref{fig:examplestwo} shows that the parallactic deviation of
the curve is also smaller than before.  It is already clear from the
lower panel that it will be difficult to extract physical information
from some of the events, particularly those with smaller Einstein
radius.

\begin{figure}
\epsfxsize=8cm \centerline{\epsfbox{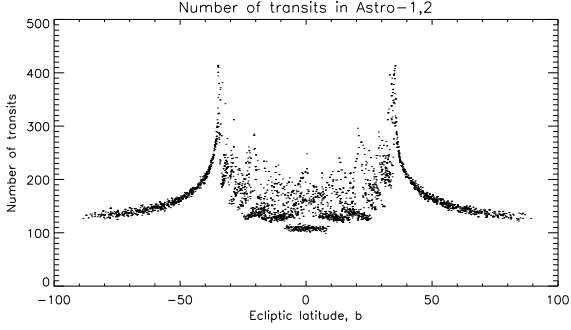}}
\caption{This shows the number of transits made by the ASTRO-1 and
ASTRO-2 telescopes as a function of ecliptic latitude. The figure
is drawn by choosing 5000 random directions on the sky, and computing
the ecliptic latitude and number of transits using software supplied
by L. Lindegren. It should be remembered that the number of transits
may be misleading, as the transits are strongly clustered into groups
of between two and five.}
\label{fig:transits}
\end{figure}

\subsection{Optical Depths}

The probability $\tau$ of detecting a microlensing event with certain
characteristics is just
\beq
\tau = \Ds \int_0^1 dx \int_0^\infty dM\, {\rho(x)\over M}
\Sigma(x,M) f(M),
\label{eq:generalprob}
\eeq
where $\Sigma$ is the area in the lens plane for which the projected
source positions yield the required effect and $f(M)$ is the mass
function or the number of lenses with mass in the range $M$ to $M+dM$.

GAIA measures displacements only along the scan, so we incorporate a
factor of $\sqrt{2}$ to take us from GAIA's one-dimensional
astrometric accuracies to the two-dimensional accuracies. The centroid
shift varies by more than $5\sqrt{2} \sigmaa$ when the projected
position of the source lies within a circle of radius $\ua \RE$, where
(Dominik \& Sahu 2000)
\beq
\ua = \sqrt{ \Tlife v \over 5\sqrt{2}\sigmaa \Dl} = \sqrt{\Tlife \thetaE \over
5\sqrt{2}\sigmaa \tE}.
\label{eq:ua}
\eeq
Here, $\Tlife$ is the lifetime of the satellite and $v$ is the
transverse source velocity in the lens plane.  Note that $5\sqrt{2}
\sigmaa$ is enough for GAIA to make a convincing measurement of the
shift, but not usually enough for detection of a microlensing event.

Instead, the minimum criterion needed to detect the event in the GAIA
database is a variation of centroid shift larger than
$5\sqrt{2}\sigmaa$ together with a closest approach between lens and
source happening with a time $\Tlife$. The area in the lens plane
giving rise to events with these characteristics is $\Sigma = 2\ua \RE
\Tlife v$.  Using (\ref{eq:generalprob}), the probability is (Dominik
\& Sahu 2000)
\beq
\taua = 4 \sqrt{G \over c^2} \Ds \langle M^{-1/2} \rangle
      \sqrt{{\Tlife^3 v^3 \over 5\sqrt{2} \sigmaa} }
       \int_0^1 dx\,\rho(x) \sqrt{1-x},
\label{eq:aopticaldepth}
\eeq
where
\beq
\langle M^{-1/2} \rangle = \int_0^\infty M^{-1/2} f (M) dM.
\eeq
We refer to the quantity (\ref{eq:aopticaldepth}) as the astrometric
microlensing optical depth.

By contrast, the probability of a microlensing event with an
amplification greater than that corresponding to the threshold of
photometric accuracy $5\sigmap$ is (e.g., Griest 1991, Dominik \& Sahu
2000), namely
\beq
\taup = {4 \pi G \over c^2} \up^2 \Ds^2
       \int_0^1 dx\,\rho(x) x(1-x),
\label{eq:popticaldepth}
\eeq
where
\beq
\up^2 = {2 \over \sqrt{1 - A_{\rm p}^{-2}}} - 2.
\label{eq:popticaldeptha}
\eeq
Here, $A_{\rm p}$ is the magnification corresponding to the threshold
photometric accuracy $5\sigmap$.  We refer to the quantity
(\ref{eq:popticaldepth}) as the photometric microlensing optical
depth. For GAIA, both the astrometric and photometric accuracies
depend on the magnitude of the source as listed in
Table~\ref{table:astroacc}.

The Einstein crossing time $\tE$ is the time taken for the source to
cross the Einstein radius. This is not the duration of a photometric
event, which is the time taken for the source to cross a circle of
radius $\up \RE$ in the lens plane. The duration of an astrometric
event is the time for which the centroid shift is greater than the
threshold $5\sqrt{2}\sigmaa$. Dominik \& Sahu (2000) show that this is
approximately the time taken to cross a circle of radius $\RE \thetaE
/ (5\sqrt{2}\sigmaa)$. So, the duration of an astrometric event is
\begin{equation}
\tae = \tE {\thetaE\over 5\sqrt{2}\sigmaa}.
\end{equation}
For GAIA, this is typically a factor of two times longer than the
duration of a photometric event.

Figure~\ref{fig:maps} show contours of astrometric and photometric
microlensing optical depth together with starcounts.  The maps assume
a standard model for the sources and lenses in the Galaxy, together
with a luminosity function and extinction law (described in Section
4).  They have been produced assuming that the relative source-lens
velocity in the lens plane is $\sim 140 \kms$ (see
Figure~\ref{fig:maybeobsa}).  Extinction is an important effect for
GAIA's microlensing capabilities, as the accuracy of both the
astrometry and photometry depends on source magnitude.  Taking into
account the effects of extinction, the all-sky averaged value of the
astrometric optical depth is $2.5 \times 10^{-5}$.  Here, the
averaging is performed by weighting the optical depth with the
starcount density. Regions like the Galactic bulge are very heavily
weighted, and so the average astrometric optical depth is of the same
order of magnitude as the typical optical depth towards the central
part of the Galaxy.  There are $\sim 10^9$ stars brighter than $V =
20$ in our model; the same is true for the Galaxy (Mihalas \& Binney
1981). This means that, during the GAIA mission, there are $\sim
25000$ astrometric microlensing events, which have a variation of the
centroid shift greater than $5\sqrt{2}\sigmaa$ together with a closest
approach during the mission lifetime.  This number can be read off
Figure~\ref{fig:maps}, by multiplying the mean astrometric optical
depth within the central contour of the upper panel by the total
number of stars within the central contour of the lower panel.  Some
of these displacements cannot be identified by GAIA as microlensing
events, first because the signal-to-noise will often be low and second
because any identification algorithm must not generate too many false
detections.  The middle panel of Figure~\ref{fig:maps} show contours
of photometric microlensing optical depth again including the effects
of extinction.  The all-sky averaged photometric optical depth is
$\sim 1.2 \times 10^{-7}$.  Let us recall that, from
eqs~(\ref{eq:popticaldepth}-\ref{eq:popticaldeptha}), a photometric
event occurs whenever there is $5\sigmap$ variation. The typical event
duration is $\sim 2$ months, so that there are a total of $\sim 3600$
photometric microlensing events during the GAIA mission.  However, a
very substantial number of these events will be undetectable.  GAIA's
sampling is sparse compared to ground-based programs like MACHO, EROS
or POINT-AGAPE, so many of these events will be missed.

\section{Error Propagation}

\subsection{Covariance Analysis}

Our first task is to understand the propagation of errors, so that we
can assess how many of the $\sim 25000$ potential astrometric
microlensing events are useful.

Since most events will not be detected photometrically by GAIA, it
will not usually be possible to combine photometric together with
astrometric data to analyze a microlensing event. The measured
quantity that will generally be provided by GAIA is the source
displacement along the scan. This contains information on the
microlensing event, but is contaminated with the source parallactic
and proper motion as well. (We assume that the GAIA datastream has
already been corrected for aberration due to satellite motion and
gravitational deflection caused by Solar System objects).

The components of ${\bf S}$ resolved with respect to right ascension
$\alpha$ and declination $\delta$ are
\begin{equation}
\left(S_{\alpha},S_{\delta}\right)= {\btheta}_{\rm s} +
\frac{{\bf u}}{u^2+2}\thetaE.
\label{eq:shiftone}
\end{equation}
The angular position of the source ${\btheta}_{\rm s}$ contains the
contributions from the source proper motion $\bmu_{\rm s}$ and source
parallax ${\bf P}_{\rm s}$. We note that
\begin{equation}
{\bf u}=\left( \begin{array}{l} 
\hat{t}\cos\phi-u_0\sin\phi-\pi_{\rm sl}P_{\alpha}\thetaE^{-1} \\  
\hat{t}\sin\phi+u_0\cos\phi-\pi_{\rm sl}P_{\delta}\thetaE^{-1} 
\end{array} \right)
\label{eq:shifttwo}
\end{equation}
where $\phi$ is the proper motion angle (Gould \& Salim 1999). We have
also expanded the parallaxes ${\bf P}_{\rm s}$ and ${\bf P}_{\rm
sl}$ as
\begin{equation}
{\bf P}_{\rm s} = \pi_{\rm s} \left( P_\alpha, P_\delta \right),\qquad
{\bf P}_{\rm sl} = \pi_{\rm sl} \left( P_\alpha, P_\delta \right),
\label{eq:shiftthree}
\end{equation}
so that $\pi_{\rm s}$ and $\pi_{\rm sl}$ are the absolute values and
$P_\alpha$ and $P_\delta$ are the direction cosines (given in van der
Kamp 1967).
 
The directions along and perpendicular to the scan are related to
right ascension and declination by a rotation, which depends on where
the satellite is pointing. As GAIA continuously scans the sky, this
angle can take any value for the same source. So, formulae
(\ref{eq:shiftone})-(\ref{eq:shiftthree}) hold good in the scan
coordinate system as well, after rotation by a random angle to take us
from the right ascension and declination to the along-scan and
across-scan coordinates. GAIA measures positions in the along-scan
direction only.

\begin{figure}
\epsfxsize=9.1cm \centerline{\epsfbox{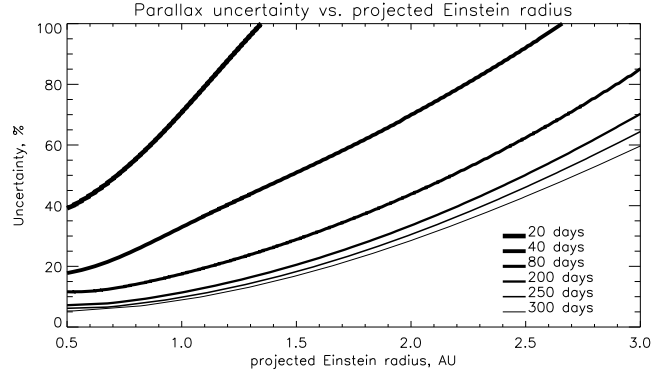}}
\caption{This shows the percentage error in estimation of the relative
parallax as a function of the projected Einstein radius $\rE$ for
different astrometric event durations and a given accuracy $\sigmaa =
150 \muas$. The events have varying source and lens distance, as well
as the transverse velocity. (All the other parameters are as in Figure
1, except $u =1$).}
\label{fig:error_rel}
\end{figure}
\begin{figure}
\epsfxsize=9.1cm \centerline{\epsfbox{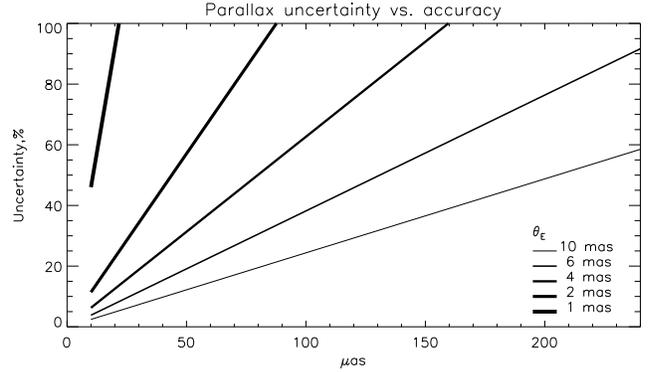}}
\caption{This shows the percentage error in estimation of the relative
parallax as a function of astrometric accuracy for different angular
Einstein radii $\thetaE$. We vary the Einstein radius by varying the
source and lens distance at fixed transverse velocity. (All the other
parameters are as in Figure 1, except $u = 5$).}
\label{fig:error_pa}
\end{figure}
\begin{figure}
\epsfxsize=9.1cm \centerline{\epsfbox{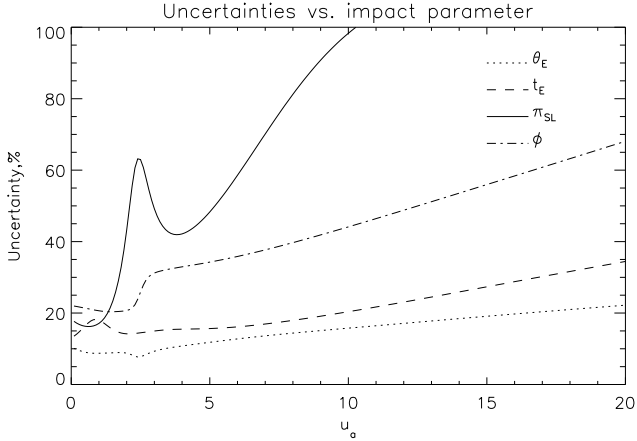}}
\caption{This shows the percentage error in estimation of the
microlensing parameters as a function of impact parameter $u$.
Accurate recovery of the relative parallax $\pi_{\rm sl}$ and the
proper motion angle $\phi$ is harder, whereas recovery of the angular
Einstein radius $\thetaE$ and the Einstein crossing time $\tE$ is
easier. (The event has the same parameters as Figure 1 with $\sigmaa =
150 \muas$.)}
\label{fig:errorprop}
\end{figure}
\begin{figure}
\epsfxsize=9.1cm \centerline{\epsfbox{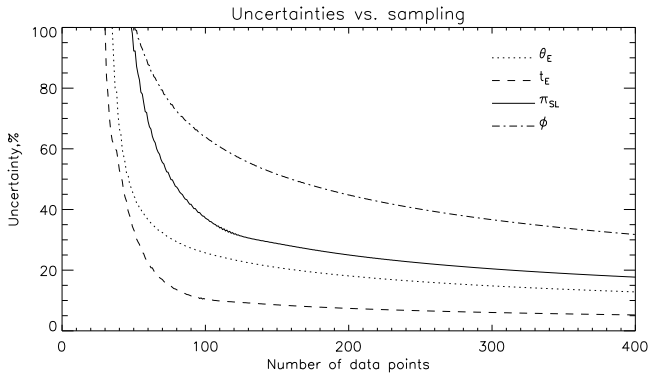}}
\caption{This shows the percentage error as a function of number of
datapoints for a typical microlensing event. Provided there are $\gta
150$ measurements, then the sampling is not the critical factor.  (The
event has the same parameters as Figure 1, but $\sigmaa = 100
\muas$).}
\label{fig:error_re}
\end{figure}

We use this microlensing model to create synthetic astrometric
measurement sets. There are a total of 11 parameters which we wish to
compute from the data, namely: $\bmu_{\rm s}$, $\pi_{\rm s}$, $u_0$,
$t_0$, $\tE$, $\phi$, $\thetaE$, $\pi_{\rm sl}$ and ${\btheta}_{{\rm
s},0}$.  To study parameter estimation performance, we evaluate the
covariance matrix $c_{ij}$ (see e.g., Boutreux \& Gould 1996; Gould \&
Salim 1999). This is defined as the matrix ${\bf C} = {\bf B}^{-1}$,
where
\begin{equation}
b_{ij}=\sum_{k=1}^N \sigmaa^{-2}\
\frac{\partial\theta_\alpha(t_k)}{\partial a_i} \frac{\partial
\theta_\alpha(t_k)}{\partial a_j}.
\end{equation}
Diagonal elements of the matrix $\sqrt{c_{ii}}$ are the individual
uncertainties in the parameters $a_i$ assuming the standard $\chi^2$
fitting procedure.  Here, $a_1,\dots,a_{11}$ are the 11 parameters
stated above and $\sigmaa$ is the astrometric accuracy for a given
star, which depends on the source magnitude and is recorded in
Table~\ref{table:astroacc}. 

Note that $t_k$ are the times at which GAIA samples the
one-dimensional astrometry datastream. In our calculations, these
times are always generated from GAIA's scanning law using Lindegren's
software.  GAIA's scanning law is quite complicated, as shown in
Figure~\ref{fig:transits}.  The number of times GAIA samples an object
$N$ is between 100 and 450, depending on its ecliptic latitude
(Lindegren 1998).  The number of transits is greatest, and so the
final astrometry is most accurate, for objects close to an ecliptic
latitude $\sim 35^\circ$. For a typical location, the number of
transits is between 100 and 200. These transits are usually clustered
into groups of between 2 and 5, so the average effective number of
independent datapoints is $\sim 40$.

\subsection{Errors in Estimates of the Astrometric Microlensing Parameters}

Although the relative parallax $\pi_{\rm sl}$ is quite hard to
estimate, it is valuable since it provides a unique solution for the
mass of the lens $M$ when combined with $\thetaE$ (see
eq.~(\ref{eq:thetae})). If in addition the source distance $\Ds$ can
be estimated -- for example, if GAIA itself measures the source
parallax -- then we have a complete solution for all the microlensing
parameters. Let us recall that in the ground-based MACHO, EROS and
OGLE programs, there is not a single microlensing event for which the
parameters can be derived unambiguously from the data
alone. Consequently, the location of the lenses responsible for the
events towards the Magellanic Clouds is a matter of considerable
controversy (e.g., Sahu 1994, Gould 1995, Alcock et al. 1997, Evans \&
Kerins 2000, Zhao \& Evans 2000). GAIA offers the promise of providing
a sample of lensing events for which all the parameters can be
extracted without any modelling assumptions.  However, some caution is
needed, as we already known in the context of the {\it Space
Interferometry Mission} (SIM) by Gould \& Salim (1999) that
astrometric measurements alone do not necessarily guarantee accurate
estimation of the relative parallax.

For which events can we recover the relative parallax to good accuracy
from the GAIA data?  First of all, an event must last an adequate time
for the microlensing shift to be distorted by parallactic movement of
the lens. Thus, the error will depend on the duration of the
astrometric event.  Second, the amplitude of the distortion is
dictated by the Einstein radius projected onto the observer plane
(e.g., Gould \& Salim 1999, Gould 2000), namely
\begin{equation}
\rE = {\thetaE \over \pi_{\rm sl}}.
\end{equation}
For close lenses ($\Dl \approx 100$ pc), it follows that $1 - \Dl/\Ds
\rightarrow 1$, and so the projected Einstein radius becomes
\begin{equation}
\rE \approx \RE \approx 1 \AU \sqrt{ {M \over \Msun}} 
\sqrt{ \Dl \over 100 {\rm \;pc}}.
\label{eq:vasilyformula}
\end{equation}
For a measurable distortion, we require $\rE$ to be about an
astronomical unit or smaller.  If it is too large, then the Earth's
motion about the Sun has a negligible effect.  Accuracy in the
relative parallax is a trade-off between duration and lens distance.

Figure~\ref{fig:error_rel} shows the uncertainty in $\pi_{\rm sl}$ as
a function of $\rE$ for different event durations $\tae$, while the
companion Figure~\ref{fig:error_pa} shows the uncertainty in $\pi_{\rm
sl}$ as a function of astrometric accuracy for different $\thetaE$. It
aids accurate recovery of the relative parallax if the angular
Einstein radius is large.  The error in $\pi_{\rm sl}$ degrades for
distant lenses. The degradation is worse for the shorter duration
events. Note that the close lenses ($\Dl \approx 100 \pc$) typically have
a crossing time $\tE \approx 1 \AU / 150$ kms${}^{-1} \approx 10$
days, and so the astrometric event duration is typically 200 days.  It
is the close lenses with longer timescales that provide the most
propitious circumstances for measuring $\pi_{\rm sl}$ from the data.

Figure~\ref{fig:errorprop} shows the uncertainties in $\thetaE,\tE,
\pi_{\rm sl}$ and $\phi$ as a function of the dimensionless impact
parameter.  A typical value of $u$ for GAIA is 11 or so.  For large
$u$, the parallactic distortion is superposed upon a small
eccentricity ellipse (see eq.~(\ref{eq:walker})). For small $u$, the
eccentricity of the ellipse becomes large and the parallactic
distortion makes the curve thinner and more elongated. This can be an
almost degenerate situation for estimating the microlensing parameters
from one-dimensional astrometry.  The percentage error in $\phi$ and
$\pi_{\rm sl}$ is larger than the error in $\thetaE$ and $\tE$.  This
is because $\pi_{\rm sl}$ controls the shape of the astrometric curve,
$\phi$ controls the orientation, while $\thetaE$ and $\tE$ the
size. Given one-dimensional astrometry, it is more difficult to
recover the details of the shape than the size.  The rise and fall in
error at $u \approx 3$ in Figure~\ref{fig:errorprop} is a consequence
of the degeneracy of the one-dimensional astrometry. Asymptotically,
the percentage error always increases with increasing impact
parameter. However, there is a r\'egime at small impact parameter when
the converse is true. The reason for this is as follows.  For very
close lenses, the parallactic distortion can become so large that the
direction of traversal of the astrometric curve is reversed. This
degeneracy is inherent in microlensing and it is not a consequence of
the limited information that can be extracted from measurements. The
astrometric curve can now correspond to events with positive impact
parameter and large parallactic effect, or negative impact parameter
and small parallactic effect. There is no way to tell the difference.

Finally, Figure~\ref{fig:error_re} shows the effects of sampling. The
error in selected microlensing parameters is plotted against the
number of datapoints for a typical microlensing event. In this Figure
only, the sampling is not performed according to GAIA's sampling law,
but the datapoints are chosen uniformly over the mission lifetime, as
we wish to illustrate the importance of sampling. In fact, GAIA samples
every object between 100 and 450 times depending on ecliptic
latitude. So, at first glance, Figure~\ref{fig:error_re} seems to
offer reassurance that sampling is not a critical factor, as it shows
that the error increases swiftly only once the effective number of
datapoints falls below 40.  However, this is misleading because GAIA's
datapoints are clustered into groups of between 2 and 5, or typically
4. This is disadvantageous, as it reduces the effective number of
datapoints by $\sim 4$, although the accuracy of the astrometry is
improved by $\sim 2$.  Figure~\ref{fig:error_re} tells us that if the
number of independent datapoints falls below $\sim 40$, then sampling
is the limiting factor in the extraction of useful parameters. (Our
calculations show that the form of the figure is only weakly dependent
on $\sigmaa$). In other words, as $\sim 40$ is roughly the average
number of effective datapoints, the sampling is responsible for
roughly tripling or quadrupling the errors in the estimated parameters
for about half the dataset.

\subsection{Errors in Estimates of Masses of Lenses}

We are primarily interested in finding the masses of the lenses. The
error in the mass is related to the errors in the angular Einstein
radius and the relative parallax by
\begin{equation}
\left( {\sigma_M \over M} \right)^2 = 4{\sigma_{\thetaE}^2 \over
\thetaE^2} + {\sigma_{\pi_{\rm sl}}^2 \over \pi_{\rm sl}} - 4 {\hat
C}(\thetaE,\pi_{\rm sl}) {\sigma_{\thetaE} \over
\thetaE}{\sigma_{\pi_{\rm sl}} \over \pi_{\rm sl}},
\label{eq:masserror}
\end{equation}
where ${\hat C}$ is the correlation coefficient between $\thetaE$
and $\pi_{\rm sl}$. Let us recall that the elements of the correlation
matrix are customarily defined as
\begin{equation}
{\hat C}(i,j) = {C_{ij} \over \sqrt{C_{ii}C_{jj}}}.
\end{equation}
The cross-term is important because errors in $\thetaE$ and $\pi_{\rm
sl}$ are strongly correlated for most of the events. This is a small
term for low impact parameter events, but it becomes important once
$\ua > 2$. The cross-term always decreases the uncertainty in the
mass. This is why Boden et al. (1998) found that eq. (14) of their paper
(which neglects the correlation) tends to overestimate the mass
error. We always use the full formula (\ref{eq:masserror}) to compute
the uncertainty in the mass in our simulations.

\begin{figure}
\epsfysize=5cm \centerline{\epsfbox{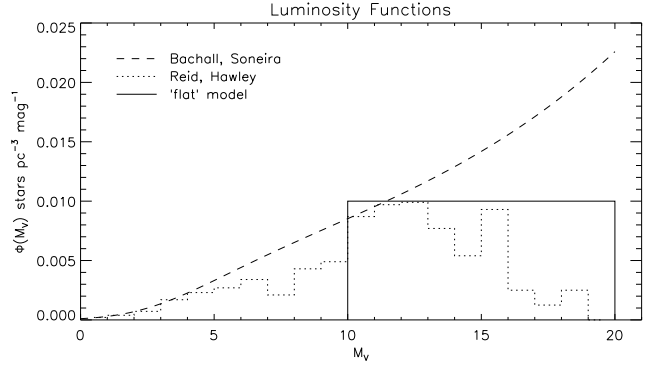}}
\caption{This shows three luminosity functions, namely the Bahcall-Soneira
(dashed line), the Reid-Hawley (dotted line) and the flat model (full
line). The flat model is the standard one in our simulations.}
\label{fig:luminosity}
\end{figure}
\begin{figure}
\epsfysize=5cm \centerline{\epsfbox{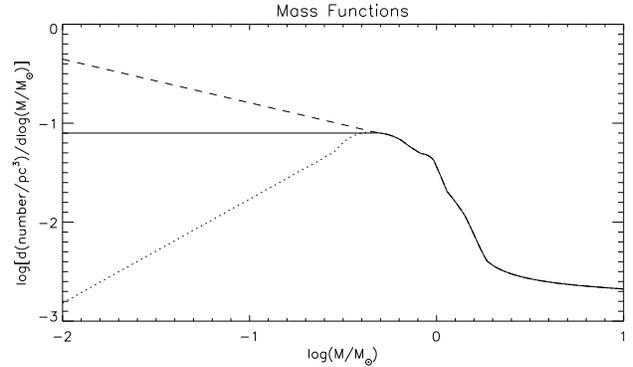}}
\caption{This shows three mass functions that are flat, rising and
falling below $0.5 \msun$. Above $0.5 \msun$, the mass function is
derived from the Reid-Hawley LF. Below $0.5 \msun$, the mass function
cannot be derived with certainty from the observations.}
\label{fig:mass}
\end{figure}

\section{A Model of the Galaxy}

The goal of the analysis of the microlensing events seen by GAIA is to
infer properties of the lenses.  The likely results can best be judged
by Monte Carlo simulations. Before this, we need to develop a model of
the lens and source populations in the Galaxy.  This section describes
the ingredients of the model in turn.

\subsection{Luminosity Function of Sources}

Let the mass density of sources be $\rho$ and the luminosity function
(LF) be $\phi$.  In an element of solid angle $d \Omega$, the number
of stars in a direction ($\ell,b$) with distances between $\Ds$ and
$\Ds + d \Ds$ and with magnitudes between $\ms$ and $\ms + d\ms$ is
\begin{equation}
dN = {1 \over \rho_0} \phi(\Ms)d\ms\rho(\ell,b,\Ds)\Ds^2 d\Ds d\Omega.
\label{eq:sourceprobability}
\end{equation}
Here, $\ell$ and $b$ are Galactic coordinates, so that the solid angle
$d \Omega$ is just $\cos b \,d\ell db$ and $\rho_0$ is the local mass
density.  As usual, absolute magnitude $\Ms$ is related to apparent
magnitude $\ms$ and extinction via
\beq
\Ms=\ms-5\log \Ds - A(\Ds, \ell,b)+5,
\eeq
where $A(\Ds,\ell,b)$ is the extinction
law. Equation~(\ref{eq:sourceprobability}) can be interpreted as a
probability density for the source parameters $\Ds, \ell, b$ and
$\Ms$. 

Three source LFs are shown in Figure~\ref{fig:luminosity}. These are
valid for the $V$ band, whereas GAIA's photometry is most accurate in
the broad $G$ band. As our aim is to provide approximate numbers in
our calculations, the conversion from $V$ to $G$ band in the LF will
not have a noticeable effect on the results.

The Bahcall-Soneira (1980) LF is a reasonably accurate predictor of
the numbers of stars in the Galaxy fainter than $G \sim 15$.  The
Reid-Hawley LF (2000) is derived from a combination of sources,
including the {\it Third Catalogue of Nearby Stars}, the {\it 8 Parsec
Sample}, as well as {\it Hipparcos} data.  It is the local LF for the
stellar disk, but it may suffer from incompleteness at magnitudes
fainter than $G \sim 16$. Accordingly, our standard assumption is that
the LF is flat in the magnitude range $10 < \Ms < 20$ and follows the
Reid-Hawley LF when $\Ms <10$.  In this case, the normalisation is
chosen to be the maximum value of the stellar luminosity function of
the Galactic disk as recorded in Table 7.3 of Reid \& Hawley
(2000). The total numbers of stars in the range $10 < G < 20$ in the
three models are given in Table~\ref{table:lf}.  The microlensing
results reported in this paper are always calculated assuming our
standard LF. This table tells us the rough scaling corrections that we
need to convert the results to other LFs.

\begin{table}
\begin{center}
\begin{tabular}{|c|c|l|}
\hline
{Number of stars} & {Number of stars} & {Luminosity
Function}\\
{(extinction)} & {(no extinction)} & \\
\hline
$0.9\times 10^9$ & $ 5.0 \times 10^9$ & flat model\\ 
$0.9\times 10^9$ & $ 5.0 \times 10^9$ & Reid, Hawley (2000)\\
$1.1\times 10^9$ & $6.1 \times 10^9$ & Bahcall, Soneira (1980)\\
\hline
\end{tabular}
\end{center}
\caption{The total number of stars with $10 < G < 20$ for
each luminosity function, including and excluding the effects
of extinction.}
\label{table:lf}
\end{table}
\subsection{Extinction}

We use the standard extinction law
\beq
A(\Ds,\ell,b) = \cases{ 0, & $|b|\!>\!50^\circ$,\cr
                        \null &\null \cr
                        {\ds 0.165(1.192\!-\!|\tan b|)
                             \over \ds |\sin b|}\times & \null \cr
                        \left[1\!-\!\exp(-{\Ds |\sin b| \over
                        h_0}) \right],& $10^\circ\!<\!|b|\!<\!50^\circ$,\cr
                        \null & \null \cr
                        \gamma \Ds, &  $0^\circ \!<\!|b|\!<\! 10^\circ$.\cr}
\eeq
For $|b| > 10^\circ$, this is the Sandage absorption law (Sandage
1972, Chen et al. 1998). The constant $h_0$ is the characteristic
height of the extinction structures, which is $\sim 120$ pc. For $|b|
< 10^\circ$, a constant differential extinction $\gamma \sim 0.5$ (in
magnitudes per kpc) is assumed. This is perhaps a little low, but we
have chosen it to ensure that the total number of stars with
magnitudes satisfying $10 \lta G \lta 20$ is $\sim 10^9$, for which
there is good evidence from starcounts (Table 4.2 of Mihalas \& Binney
1981).  Our extinction model is a reasonably accurate local
description, but it does not include features like the molecular ring
at $\sim 4$ kpc. Nonetheless, it will be good enough for our study, as
the sources are generally nearby.

In the Monte Carlo simulations, once the source distance, direction
and absolute magnitude are chosen, then the extinction law is used to
work out the apparent magnitude.  We calculate the astrometric
accuracy $\sigmaa$ for a given source magnitude using the
accuracy-magnitude relation for GAIA, listed in Table~\ref{table:astroacc}.

\subsection{Density of Sources}

The density of sources at a distance $\Ds$ in the direction of
Galactic longitude $\ell$ and latitude $b$ is
\beq
\rho=\rho_{\rm d} + \rho_{\rm b}.
\eeq
Here, $\rho_{\rm d}$ is the density distribution of the Galactic disk
and $\rho_{\rm b}$ is the density of the Galactic bulge. We use an
exponential disk model with scalelength $\Rd \sim 3$ kpc and
scaleheight $\zh \sim 350$ pc (Gould, Bahcall \& Flynn 1997), namely
\beq
\rho_{\rm d}=\rho_0 \exp\left(-(R-R_0)/\Rd- |z|/\zh\right),
\label{eq:expd}
\eeq
where $R$ is Galactocentric distance, $z$ is the height above or below
the Galactic plane, $R_0$ is the radial position of the Sun and
$\rho_0$ is the local density of disk stars, which is taken as $0.10
\Msun \pc^{-3}$ (Holmberg \& Flynn 2000).  It is straightforward to
establish that
\beq
R^2=R_0^2\!+\!\Ds^2\cos^2 b\!-\!2\Ds R_0 \cos b \cos\ell,\quad
z=D_s \sin b.
\eeq
We use the oblate axisymmetric bulge model provided by Kent (1992)
\begin{equation}
\rho_{\rm b}(R,z) = 3.53\, K_0\left({s\over 667\,\pc}\right) \,\msun \pc^{-3}, 
\end{equation}
where $s^4 = R^4+(z/0.61)^4$ and $K_0$ is a modified Bessel
function. Kent's model is a reasonable fit to the Spacelab infrared
data. However, it is known to be wrong in detail as the Galactic bulge
is really a triaxial bar (e.g., Binney et al. 1991; H\"afner et
al. 2000). This means that our model -- like any axisymmetric model --
will underestimate the photometric optical depth as compared to the
observations in windows close to the Galactic Centre typically by a
factor 2 or so (see e.g., Evans 1994, 1995; Binney 2000).

\subsection{Mass Function and Density of Lenses}

It is nearby objects that make the largest contribution to the
astrometric lensing signal.  Accordingly, in our Monte Carlo
simulations, only the disk stars act as lenses. Halo populations of
dark objects like Machos are not included.  So, the density of lenses
is the double exponential disk prescribed by eq~(\ref{eq:expd}).  We
generate the masses of the lenses using the composite mass function
shown in Figure~\ref{fig:mass}.  Above $0.5 \msun$, we use a mass
function (MF) derived from the Reid-Hawley LF via
\beq
f (M) = \phi( \Ms (M)) {d \Ms \over dM},
\label{eq:masslum}
\eeq
Here, $\Ms (M)$ is the relationship between mass and absolute
magnitude, as recorded in Kroupa, Tout \& Gilmore (1990).  This
ensures that the high-mass end of the MF (which gives most of the
light) is compatible with the local LF. The low-mass end of the MF is
not at all certain and its derivation from the LF is fraught with
difficulties (e.g., D'Antona \& Mazzitelli 1994). Hence, for masses
below $0.5 \msun$, we assume that the MF is flat in our standard
model.

\subsection{Velocity Distributions}

We draw components of source and lens velocities from triaxial Gaussian
distributions:
\begin{eqnarray}
f(V_R,V_z,V_{\phi})&=&\frac{1}{(2\pi)^{3/2} \sigma_R\sigma_\phi\sigma_z} \nonumber\\
&\times& \exp(-\frac{V_R^2}{2\sigma_R^2}
- \frac{(V_{\phi}-\vmean)^2}{2\sigma_\phi^2}
-\frac{V_z^2}{2\sigma_z^2})
\label{eq:velocityprobability}
\end{eqnarray}
For the bulge sources, the random velocity is $\sigma_R = \sigma_\phi=
\sigma_z = 100 \kms$ about a mean of $50 \kms$ (McGinn et al. 1989).
For the disk sources, the random component has $\sigma_R = 34 \kms$
$\sigma_\phi = 21 \kms$ $\sigma_z = 18 \kms$ about a mean velocity
$\vmean$ of $214 \kms$ (Edvardsson et al. 1993). The assumption of
triaxial Gaussians with constant semiaxes seems reasonable and is
often made in studies of Galactic microlensing (e.g., Kiraga \&
Paczy\'nski 1994, Kerins et al. 2001). Nonetheless, a more exact
calculation should derive the spatial variation in the means and
dispersions of the velocities of the stellar populations by solving
the Jeans equations.

\begin{figure*}
\epsfysize=15.cm \centerline{\epsfbox{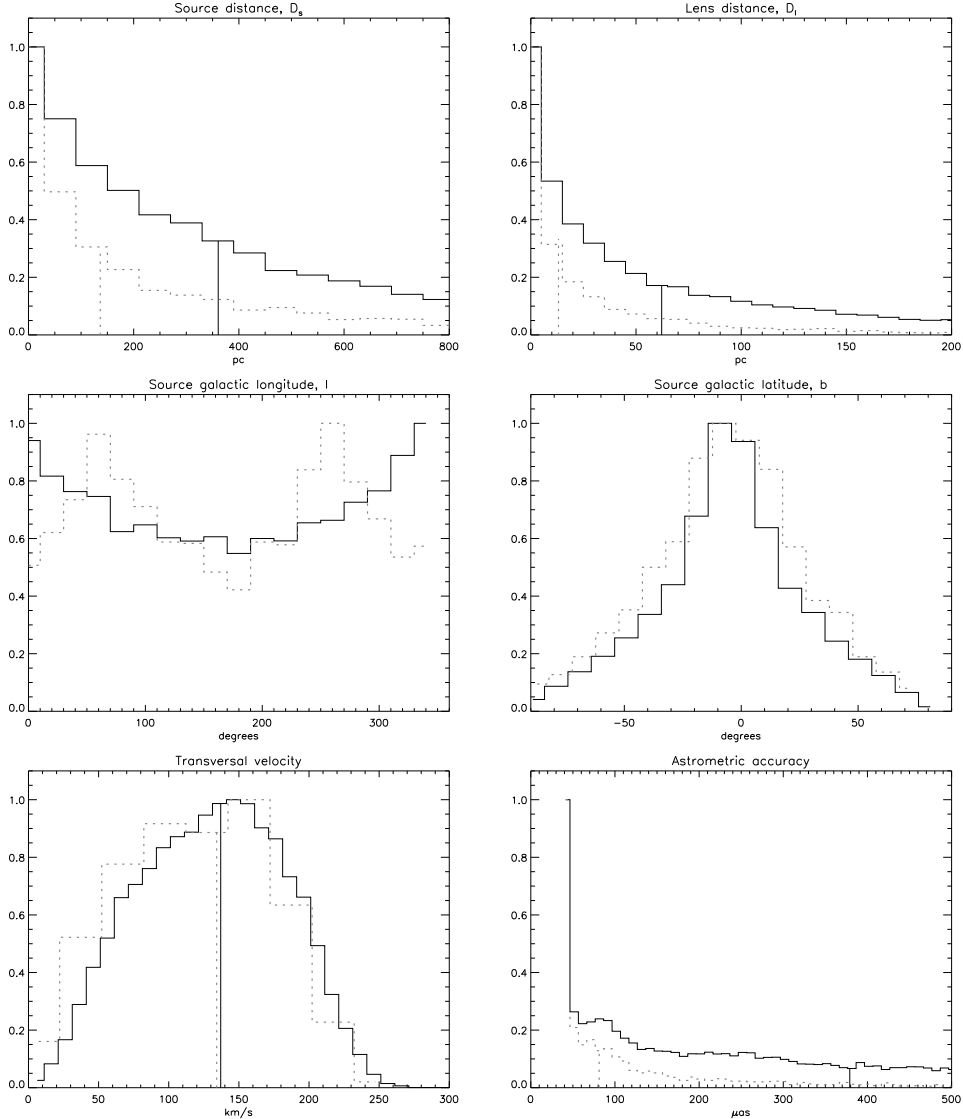}}
\caption{Distributions of source distance $\Ds$, lens distance $\Dl$,
source coordinates $(\ell, b)$, transverse velocity $v$ and
astrometric accuracy $\sigmaa$ for the simulated datasets.  The
histogram in full lines refers to the whole sample, dotted lines to
the sub-sample of high quality events. The distributions are all
normalised to a maximum value of unity. The medians are indicated by
vertical bars.}
\label{fig:maybeobsa}
\end{figure*}
\begin{figure*}
\epsfysize=15.cm \centerline{\epsfbox{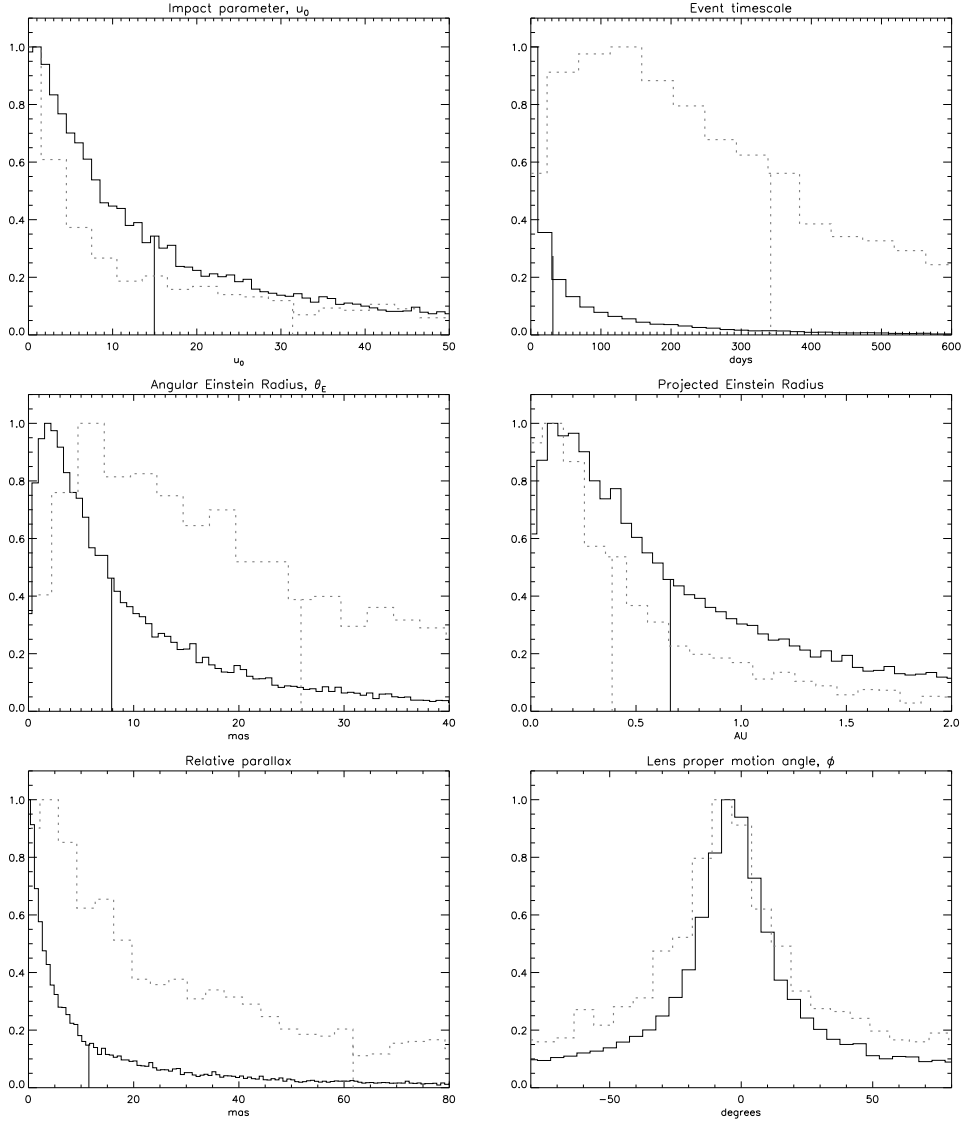}}
\caption{Distributions of impact parameter $\ua$, event duration
$\tae$, angular Einstein radius $\thetaE$, projected Einstein radius
$\rE$, relative parallax $\pi_{\rm sl}$ and lens proper motion angle
$\phi$ for the simulated datasets.  The histogram in full lines refers
to the whole sample, dotted lines to the sub-sample of high quality
events. The distributions are all normalised to a maximum value of
unity. The medians are indicated by vertical bars.}
\label{fig:maybeobsb}
\end{figure*}

\begin{table}
\begin{center}
\begin{tabular}{|c|c|c|}
\hline
{Parameter} & {Whole Sample} & {High Quality Sample} \\
\hline
$\tae$ (days) & 32 (163) & 342 (814)\\ 
$\thetaE$ (mas) & 7.9 (20.0) & 26.0 (62.3)\\
$\rE$ (au) & 0.7 (1.3) & 0.4 (0.7) \\
$\pi_{\rm sl}$ (mas) & 11.5 (244) & 62.0 (1066)\\
$|u_0|$ & 15 (51) & 31 (140)\\
$\Dl$ (pc) & 62 (209) & 14 (44)\\
$\Ds$ (pc) & 360 (800) & 137 (326)\\
$v$ ($\kms$)& 135 (137) & 132 (134)  \\
$\sigmaa$ ($\tmuas$) & 380 (570) & 81 (171) \\
$G {\rm \;magnitude}$ & $\approx$ 18 (18) & $\approx$ 15 (16) \\
\hline
\end{tabular}
\end{center}
\caption{This records the median and the mean (in brackets) values of
the parameters of the simulated events for the whole sample and for
the subset of high quality events (those with mass uncertainty less
than $50 \%$).}
\label{table:medians}
\end{table}
\section{Monte Carlo Simulations}

We first pick the source position, velocity and magnitude, which
provides the astrometric accuracy.  This is done by drawing randomly
from the probability distributions implied by
eqs.~(\ref{eq:sourceprobability}) and (\ref{eq:velocityprobability}),
using the rejection method (e.g., Press et al. 1992).  The astrometric
optical depth (\ref{eq:aopticaldepth}) can be recast to provide the
joint probability of drawing the lens distance and transverse velocity
for a given source. We draw the mass of the lens from the mass
function.  Finally, we pick the impact parameter $u_0$ and the phase
$t_0$ by drawing uniformly from the rectangular area in the lens plane
of size $\Tlife v$ and $2\ua\RE$ (illustrated in Figure 2 of Dominik
\& Sahu (2000)). This provides us with a single simulated microlensing
event, which has a variation of the centroid shift at least as large
as $5\sqrt{2}\sigmaa$ and which has a maximum during the lifetime of
the GAIA mission.

\subsection{Distributions of the Events}

We create a set of 25000 such microlensing events, as a synthetic GAIA
database.  For each member of the sample, we compute the error in the
estimated mass of the lens using eq~(\ref{eq:masserror}), which of
course depends on the uncertainties in the relative parallax and the
angular Einstein radius. From this, we build up a subset of 2531
events for which the error in the inferred mass is less than $50
\%$. We refer to this as the sample of high quality
events. Figures~\ref{fig:maybeobsa} and~\ref{fig:maybeobsb} show the
distributions of the microlensing parameters for the entire sample and
for the subset of high quality events. The medians and means of the
distributions are given in Table~\ref{table:medians}. Note that the
distributions have long tails caused by high impact parameter events
and so we typically use medians rather than means, in accord with
normal statistical practice.

There are a number of interesting things to notice.  The mean source
distance is $\sim 800$ pc for the whole sample, and about 140
pc for the high quality events. The mean lens distances are still
smaller. So, the astrometric microlensing signal seen by GAIA is
overwhelmingly dominated by the very local stars. Of course, more
distant microlensing events take place, but they can be much less
commonly identified and still less commonly analysed to extract the
parameters for the event.

For the whole sample, the optimum direction is towards the Galactic
Centre. This is not surprising, as the source and lens density is
highest in this direction. However, the high quality events are found
preferentially at longitudes $\ell \sim 70^\circ$ and $260^\circ$.
These events have longer durations than usual. The largest component
of the transverse velocity is the circular motion about the centre of
the Galaxy. So, it helps to be looking in directions in which the
circular velocity is almost entirely along the line of sight. Our
model of the Galaxy is axisymmetric, so the peaks in the dashed
histogram of longitude are roughly symmetric about $\ell =0^\circ$.  A
similar effect can be seen in the distribution of Galactic
latitude. Here, the high quality events have a broader distribution
than the whole sample. A longer duration event is favoured by a
somewhat higher latitude as the velocity component transverse to the
line of sight is diminished.

In understanding the distribution of accuracies $\sigmaa$, we recall
that the events in the whole sample have been chosen to have at least
a $5\sqrt{2}\sigmaa$ variation in the centroid shift. If the source is
faint, then the accuracy of the astrometry is poor and very few events
can satisfy this stringent criterion. We recall from eq~(\ref{eq:ua})
that $\sigmaa$ is proportional to $\thetaE$ and varies inversely with
$u^2$. So, the high quality sample is still more strongly biased to
nearby luminous stars for which $\sigmaa$ is small.

There are two ways to obtain a high quality event. Either the
astrometric accuracy is good or the microlensing event has optimum
characteristics, namely that the angular Einstein radius is large, the
duration is long and the projected Einstein radius $\rE$ is comparable
to the size of the Earth's orbit around the Sun. All these
characteristics are evident on comparing the full and the dashed
histograms in Figure~\ref{fig:maybeobsb}.  The median duration of the
high quality events is $\sim 350$ days, nearly a factor 10 times
larger than the whole sample. The projected Einstein radius $\rE$
already has a median of 0.7 au for the whole sample, and it is still
smaller at 0.4 au for the high quality events. The median angular
Einstein radius is nearly a factor of 4 larger for the high quality
events. Note that the median impact parameter for the high quality
events is larger than that for the whole sample. This is because $u_0
\sim \sqrt{\thetaE / \sigmaa}$. For a high quality event, it helps to
have large angular Einstein radius and good accuracy, so the
distribution of impact parameters has a long tail.

Table~\ref{table:tableoferrors} shows the percentage of events for
which $M, \tE, \thetaE$ and $\pi_{\rm sl}$ can be recovered to within
$50 \%$. From the table, we see that over $40 \%$ of all the events
will have good estimates for $\tE$ and $\thetaE$ from the GAIA
astrometry alone.  Current ground-based programs like MACHO and EROS
attempt to recover the characteristic masses of lenses from estimates
of $\tE$ alone for samples of a few tens to hundreds of events. By
contrast, GAIA will provide a much larger dataset of $\sim 10000$
microlensing events with good estimates of both $\tE$ and
$\thetaE$. The typical locations and velocities of the sources and
lenses can be inferred for the ensemble using statistical techniques
based on Galactic models (in much the same manner as MACHO and EROS do
at present).

The number of events with an amplification $A$ corresponding to $5
\sigmap$ and with at least one datapoint sampled on the photometric
lightcurve by GAIA is $1260$.  So, $\sim 5\%$ of all the $25000$
events have a photometric signal recorded by GAIA. One datapoint is of
some help as it provides corroborating evidence that a microlensing
event has taken place. If we ask for at least nine datapoints (which
may very well be clustered, so there may be only three independent
measurements), then the number of events is $427$. So, for only $\sim
2\%$ of all the astrometric events will GAIA itself obtain photometry
which will substantially improve the parameter estimation.

\begin{table}
\begin{center}
\begin{tabular}{|c|c|}
\hline
\null& Percentage \\ \hline
$N(\sigma_M < 50 \%)$ & 10 \%\\
$N(\sigma_{\tE} < 50 \%)$ & 37 \%\\
$N(\sigma_{\thetaE} < 50 \%)$ & 46 \%\\
$N(\sigma_{\pi_{\rm sl}} < 50 \%)$ & 13 \%\\
\hline
\end{tabular}
\end{center}
\caption{This gives the percentage of events in the whole sample with
an error of less than $50 \%$ in the estimate of mass $M$, or the
Einstein crossing time $\tE$, or the angular Einstein radius
$\thetaE$ or the relative parallax $\pi_{\rm sl}$.}
\label{table:tableoferrors}
\end{table}
\begin{figure}
\epsfysize=5.cm \epsfxsize = 8.5cm \centerline{\epsfbox{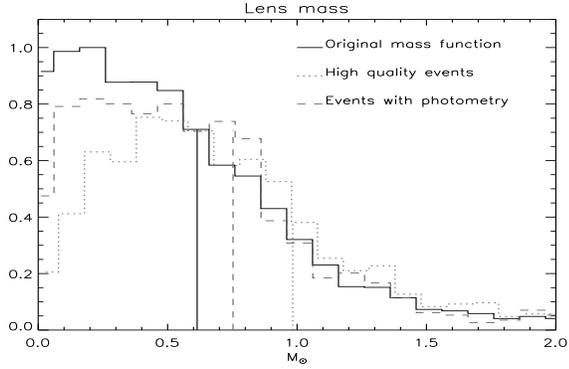}}
\caption{The distribution of masses of the lenses for the whole sample
(full histogram), the subsample of high quality events (dotted
histogram) and the subsample of events with some photometry (dashed
histogram). The bars indicate the median of the distributions. Notice
the bias at low masses in the high quality sample.}
\label{fig:massbias}
\end{figure}
\begin{figure}
\epsfxsize=8.5cm \centerline{\epsfbox{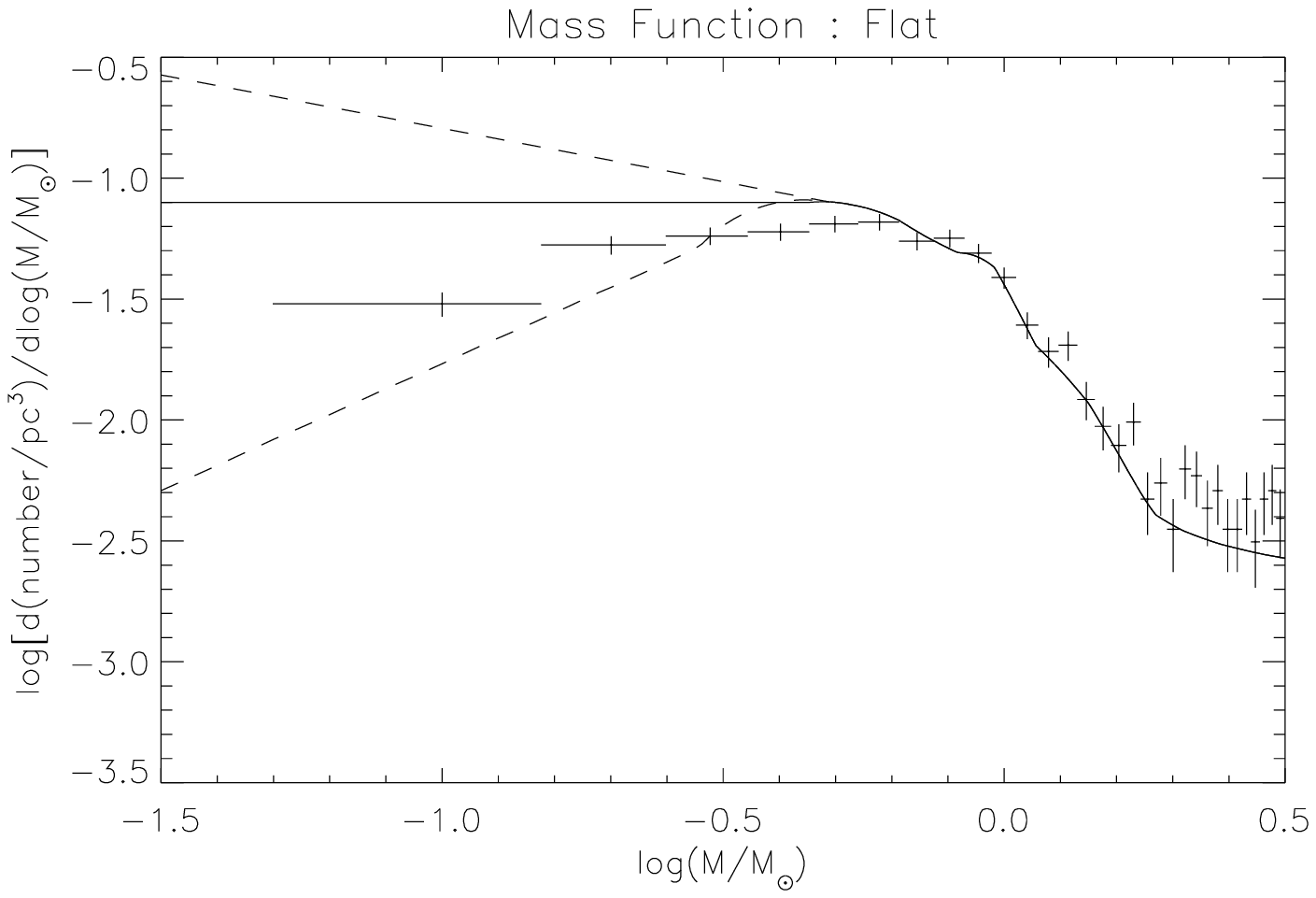}}
\epsfxsize=8.5cm \centerline{\epsfbox{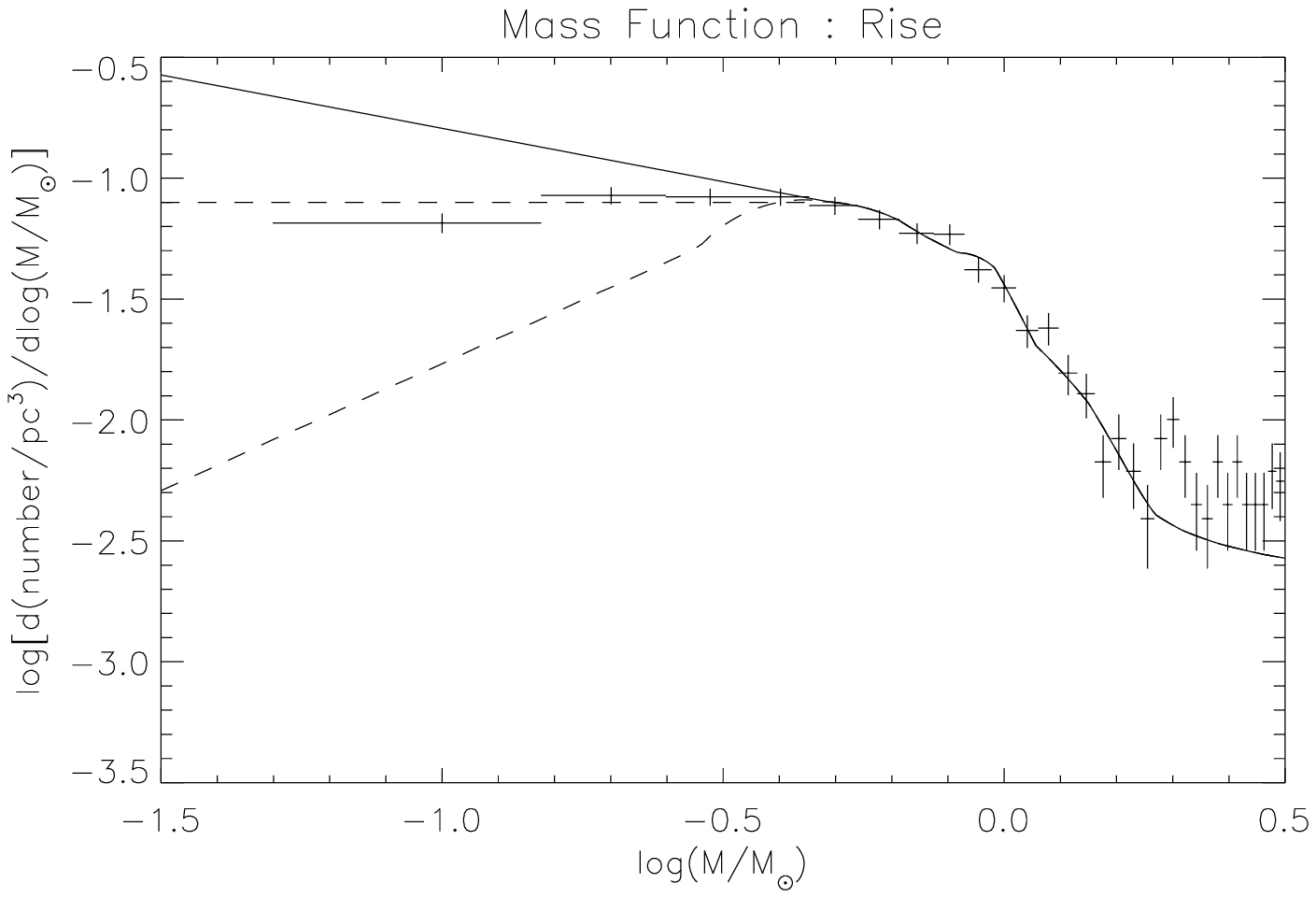}}
\epsfxsize=8.5cm \centerline{\epsfbox{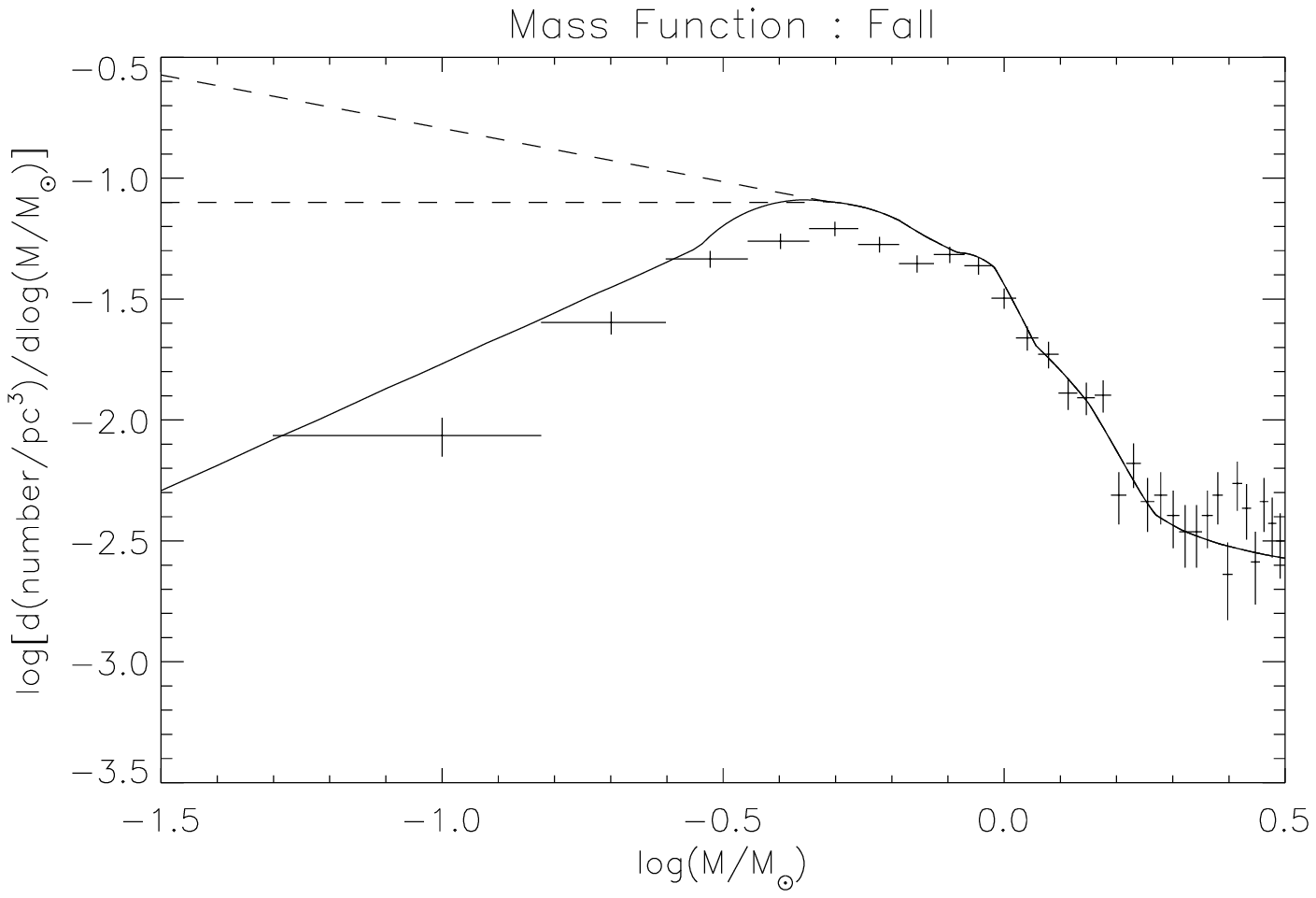}}
\caption{The recovery of the flat, rising and falling mass functions
from the subsamples of high quality astrometric microlensing events
generated from simulations.  The vertical error bar associated with
each bin is proportional to the square root of events in the bin.  The
horizontal error bar is the size of the bin. 
}
\label{fig:massfuns}
\end{figure}

\subsection{Determination of the Local Mass Function}

GAIA is the first instrument to have the capabilities of detecting
nearby populations of very dark objects. For example, local
populations of low mass ($\sim 0.5 \msun$) black holes could easily
have eluded identification thus far. Cool halo white dwarfs are also
extremely difficult to detect. Existing programs look for faint
objects with high proper motions characteristic of halo populations
(e.g., Oppenheimer et al. 2001). Such programs miss stars with low
space motions and have difficulties detecting stars with extremely
high proper motions (depending on the epoch difference of the
plates). So, they can only set lower limits to the local halo white
dwarf density.  Samples of disk white dwarfs are again selected
kinematically and are subject to similar biases (Knox, Hawkins \&
Hambly 1999). There may remain undetected a larger population of even
fainter and cooler white dwarfs, as it is unclear whether existing
surveys are probing the very faintest luminosities
possible. Similarly, there may be an extensive local population of
very cool brown dwarfs. The {\it 2 Micron All Sky Survey} (2MASS) and
the {\it Sloan Digital Sky Survey} (SDSS) have increased the known
population of local dwarf stars with spectral types later than M to
over a hundred (e.g., Strauss et al. 1999, Burgasser et al. 2000,
Gizis, Kirkpatrick \& Wilson 2001).  Nonetheless, the coolest brown
dwarfs will have easily eluded the grasp of these and all current
surveys. Future projects like the {\it Space Infrared Telescope
Facility} (SIRTF) may be able to detect very old brown dwarfs in the
solar neighbourhood by using long integration times, but a
large-scale survey will be prohibitively costly in terms of time.
However, the astrometric microlensing signal seen by GAIA will be
sensitive to local populations of even the dimmest of stars and
darkest of these objects. 

Here, we investigate whether we can recover the local mass function
(MF) from the astrometric microlensing signal seen by GAIA. We begin
by examining the selection effects in the subsample of high quality
events, the only ones for which individual lens masses can be
recovered to reasonable accuracy.  Figure~\ref{fig:massbias} shows the
distribution of masses for the entire sample and for the high quality
events. The latter are biased towards masses greater than $0.3
\msun$. This is not too surprising, as it is more difficult to recover
the relative parallax if the Einstein radius is small. If instead of
the high quality events, we consider the subsample of events with at
least one corroborating photometry datapoint, then the bias is
reduced.  Figure~\ref{fig:massbias} already suggests that GAIA's
astrometric microlensing signal will be an unbiased and powerful probe
of the local population of stellar remnants, such as white dwarfs and
neutron stars. However, the signal seen on smaller mass-scales will
required modelling and correction for selection effects before it is
unbiased.

Reid \& Hawley (2000) compute the mass function (MF) for nearby stars
from the volume-limited {\it 8 Parsec Sample} using empirical
mass-luminosity relationships. They obtain a power-law MF ($f(M)
\propto M^{-1}$) over the range $0.1$ to $1.0 \msun$. This is in
rough agreement with the MF as inferred from studies of the LF of red
disk stars seen with {\it Hubble Space Telescope} (Gould, Bahcall,
Flynn 1997), which is a broken power-law. Above $\sim 0.6 \msun$, the
MF is nearly of the classical Salpeter form with $f(M) \propto
M^{-2.21}$. Below $0.6 \msun$, it is flatter than Salpeter with $f(M)
\propto M^{-0.9}$ (after correction for binaries).  Of course, the
behaviour of the MF becomes more uncertain as the hydrogen-burning
limit is approached.

We use the three MFs shown in Figure~\ref{fig:mass} as spanning a
range of reasonable possibilities. Above $0.5 \msun$, the MF is always
derived from the Reid-Hawley luminosity function. Below $0.5 \msun$,
there are three possibilities. It may be flat ($f(M) \propto M^{-1}$),
rising ($f(M)\propto M^{-1.44}$) or falling ($f(M) \propto
M^{0.05}$). These choices are motivated by the Reid-Hawley LF as shown
in Figure~\ref{fig:luminosity}. For the falling MF, we choose the
power-law index to be that implied by the decline in the LF below $\Ms
\approx 13$. For the rising MF, we choose the power-law index by
assuming that the rise at $\Ms \approx 11$ continues to fainter
magnitudes and that these stars are missing from the data because of
incompleteness. Of course, the flat MF is so called because it has
equal numbers of objects in each decade of mass and so appears flat in
a plot of $\log Mf(M)$ versus $\log M$.  For each of the three MFs, we
generate samples of 25000 astrometric microlensing events using Monte
Carlo simulations. We extract the high quality events and compute the
mass uncertainty using the covariance analysis. (In actual practice,
the high quality events would be selected on the basis of the goodness
of their $\chi^2$ fits).  We build up the MF as a histogram.  The
vertical error bar associated with each bin is proportional to the
square root of events in the bin.  The horizontal error bar is the
size of the bin.

The three cases are shown in Figure~\ref{fig:massfuns} with the solid
line representing the underlying MF. The simulated datapoints with
error bars show the MFs reconstructed from the high quality events.
It is evident that GAIA can easily distinguish between the flat,
rising and falling MFs. The MFs are reproduced accurately above $\sim
0.3 \msun$. Below this value, the reconstructed MFs fall below the
true curves, as a consequence of the bias against smaller Einstein
radii. However, this does not compromise GAIA's ability to
discriminate between the three possibilities. In practice, of course,
simulations could be used to re-calibrate the derived MFs at low
masses and correct for the bias.

We have also carried out simulations with MFs containing spikes of
compact objects, such as populations of $\sim 0.5 \msun$ white dwarfs.
They lie in the mass r\'egime to which GAIA's astrometric microlensing
signal is most sensitive. So, such spikes stand out very clearly in
the reconstructed MFs.  We conclude that one of the major scientific
contributions that GAIA can make is to determine the local MF.
Microlensing provides the only way of measuring the masses of
individual objects irrespective of their luminosity.  GAIA is quite
simply the best survey instrument to carry out an inventory of masses
in the solar neighbourhood.
\begin{figure}
\epsfysize=10cm \centerline{\epsfbox{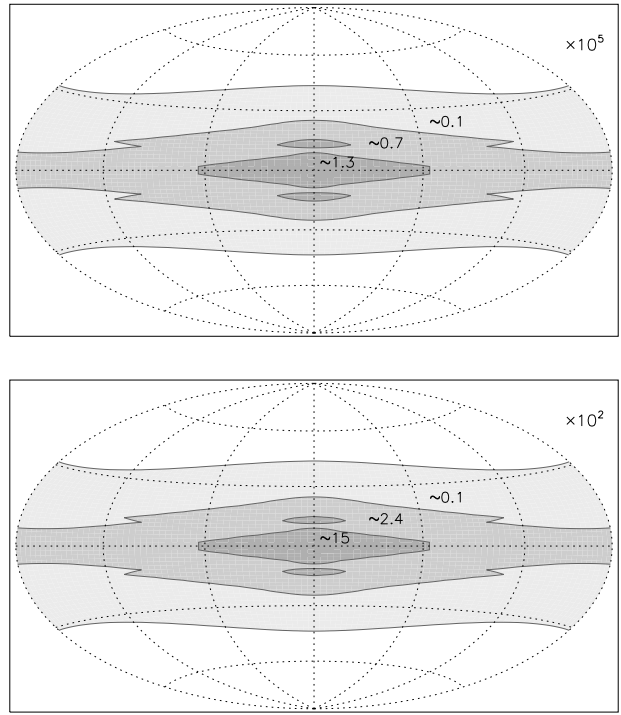}}
\caption{This shows contours of ``microlensing noise'' in astrometry
(upper panel) and photometry (lower panel).  The number within each
contour refers to the total number of stars inside the shaded region
enclosed between the contours that suffer an astrometric distortion or
photometric brightening exceeding the mission target accuracy. The
numbers are in units of $10^5$ in the upper panel and $10^2$ in the
lower panel.}
\label{fig:noisy}
\end{figure}
\begin{figure}
\epsfxsize=9.1cm \centerline{\epsfbox{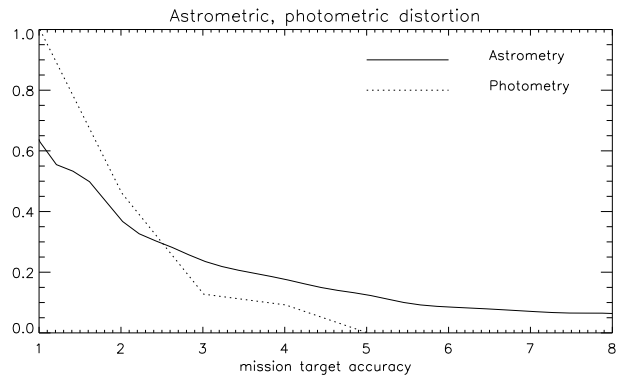}}
\vskip 0.5truecm
\caption{This shows the distribution of deviations due to microlensing
at the midpoint of the mission in units of the astrometric and
photometric mission target accuracy.  The distributions are all
normalised to a maximum value of unity.  (For agiven source, the
mission target accuracy is that achieved at the end of the GAIA
mission. For astrometry, this is better than the instantaneous
accuracy listed in Table~\ref{table:astroacc} by a factor of $\sim
10$).}
\label{fig:noisydist}
\end{figure}

\section{Strategy}

\subsection{Detection of Events}

All sources detected by GAIA will be fit using the standard algorithms
for the 6 astrometric parameters, namely ${\btheta}_{{\rm s},0}$,
${\bmu}_{\rm s}$ and ${\bf P}_{\rm s}$.  This is described in, for
example, volumes 1 and 3 of {\it The Hipparcos and Tycho Catalogues}
(ESA 1997).  So, the first cut on the dataset is to ask for large
residuals in the standard astrometric parameters. Of course,
microlensing is just one cause of large residuals out of many. Other
possibilities include multiple companions, eclipsing systems, the
mottled or spotted surfaces of stars and variability-induced movers or
VIMs (Wielen 1996). VIMs were first discovered in {\it Hipparcos}
data, as they give spurious negative parallaxes when fit to the
standard astrometric model.  In VIMs, it is hypothesised that the
photometric centre of a close double star moves on the sky as one of
the components has variable luminosity.

Suppose we detect the event photometrically as well as
astrometrically.  Then, GAIA's spectroscopy and multicolour photometry
can help us distinguish microlensing events from other possible causes
of astrometric shifts.  GAIA's spectroscopy is reasonably accurate for
objects brighter than $V \sim 15$, although it covers a short
wavelength range near the Ca triplet. For objects fainter than $V \sim
15$, GAIA's mulicolour photometry will be more useful.  If the lens is
dark, then the colours before, during and after the microlensing event
are the same. Such is not true of spotted stars, VIMs and resolved
binaries, which will show colour changes over the course of time. If
the lens is luminous, then the colour does change over the course of
the microlensing event. However, GAIA will detect such a luminous lens
and so the lens proper motion and parallax will be available from a
standard astrometric fit. This means that we can correct for the
gravitational deflection of the source directly from the data. The
multicolour photometry can also help us reduce the false detection
rate.  An unblended microlensing event is achromatic and does not
repeat, whereas most other types of variability are chromatic and do
repeat. Even long-period variables have periods of at most $\sim 600$
days (Cox 2000) and thus are short enough for GAIA to recognise as
repeating.

In propitious circumstances, close companions can be distinguished
from microlensing by analysis of the rate of change of the astrometric
excursion with time. At times far from the closest approach, the
source motion is slow in a microlensing event, whereas at or near the
closest approach, the source rapidly traverses the perturbation
ellipse. Although a nearby companion also causes an elliptical
distortion to be superposed on the proper and parallactic motion, the
traversal is harmonic in time and repeats. A close companion will also
induce radial velocity variations, which may be detectable with GAIA's
spectroscopy. If the stars in the binary are of similar luminosity but
different colour, then there may be color variations during the
centroid motion as well.

If the closest approach, which corresponds to the maximum astrometric
deviation, does not occur during the lifetime of the GAIA mission,
then it will be difficult to be certain that the centroid shift is
caused by microlensing. This is why, in this paper, we have restricted
attention only to events which satisfy the stringent criteria of a
$5\sqrt{2}\sigmaa$ variation in the centroid shift and a maximum
during the mission lifetime. 

\begin{table*}
\begin{center}
\begin{tabular}{c|cccc}\hline
$u_0$ & 0.25 & 0.5 & 1 & 5 \\
$\%$ error & 0.28 & 0.26 & 0.22 & 0.14 \\
\hline
\end{tabular}
\end{center}
\caption{This table lists the percentage error in the estimated 
point spread function against the impact parameter of the microlensing
event.}
\label{table:psferror}
\end{table*}

\subsection{Noise caused by Microlensing}

``Microlensing noise'' is the name given to the weak effects of
typically large impact parameter microlensing events. This can cause
fluctuations in the positions of stars. There is some analogy with the
effects of turbulence in the upper atmosphere, which causes wavefronts
to become corrugated and hence causes sources to blur and
twinkle. Microlensing noise is a fundamental limit to the accuracy to
which we can know the positions of sources.  For example, Sazhin et
al. (1998, 2001) have pointed out that fluctuations of the angular
positions of reference extragalactic radio and optical sources will be
caused by the microlensing effects of stars in the Galaxy.  Such
angular fluctuations may range up to hundreds of microarcseconds.
Sazhin et al. (1998) found a good example of a quasar with large
positional uncertainties caused by the lensing effects of a nearby
{\it Hipparcos} star; the distance of the star is very close, just 50
pcs.

Figure~\ref{fig:noisy} shows a contour plot of number of sources which
have an astrometric or photometric variation caused by microlensing of
the same order of magnitude as the mission target accuracy. (Let us
recall the mission target accuracy is that achieved at the end of the
mission and so is the instantaneous astrometric accuracy given in the
final line of Table~\ref{table:astroacc} divided by $\sim 10$).  The
total number of sources having an astrometric deviation exceeding the
mission target accuracy is $2.2 \times 10^5$.  The positional
measurement of one source in every twenty thousand is affected by
microlensing noise at any instant. This number is obtained by
computing $(2.2 \times 10^5 / 10^9) \times \langle \tae \rangle / 5\;
{\rm years}$. Many of these events will be long duration events. Any
astrometric deviation will often be only slowly varying over the
mission lifetime and so this will not compromise the proper motion and
parallax target accuracies for GAIA.  For photometry, the
instantaneous number of stars affected by microlensing noise is $\sim
1.8 \times 10^3$.  Figure~\ref{fig:noisydist} shows the distribution
of these disturbances in units of the mission target accuracy. To be
included on this Figure, an event must already have a deviation
greater than the mission target accuracy. It is reassuring that the
mean of the distribution of photometric deviations is strongly peaked
near unity, so the blurring in photometry is mostly insignificant.

If the distribution of errors caused by instrumental effects is
Gaussian, then the number of events with error less than $0.1 \sigmaa$
is just $\sim 10 \%$.  With the more realistic assumption of a
non-Gaussian error distribution with higher wings the number of such
deviation will be still smaller.  So, if we assume that microlensing
noise is a random error source, then its effects for GAIA are
negligible, as in $\sim 90 \%$ of cases the microlensing noise is less
than that expected from instrumental sources.

Another way in which the microlensing noise manifests itself is in
distortion of the point spread function (PSF). GAIA's PSF may be
approximated as a biaxial Gaussian with FWHM of 300 mas and 150 mas.
To investigate the likely size of the errors, we convolve the PSF with
the two micro-images corresponding to an event with Einstein radius
$\thetaE = 5$ mas. These micro-images are taken as pure
$\delta$-functions.  The convolved PSF is then fit with the standard
biaxial Gaussian. This procedure of course carries an error, which we
estimate by computing the area difference between the true PSF and
fit. This is divided by the area under the PSF to give the fractional
error caused by microlensing. Table~\ref{table:psferror} list the
percentage error as a function of impact parameter. Although the error
is small, it will be important as GAIA is aiming for a photometric
accuracy of fractions of a percent.

\begin{figure}
\epsfxsize=8.5cm \centerline{\epsfbox{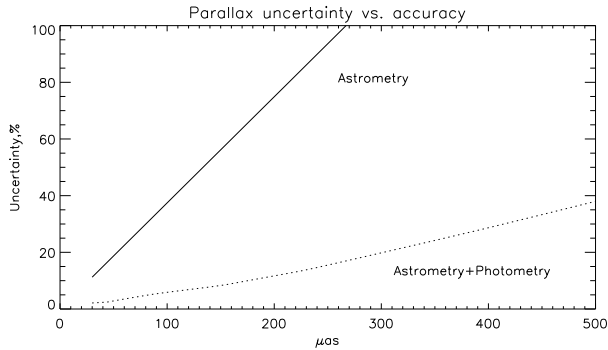}}
\caption{This shows the percentage error in the parallax of the lens
as a function of the astrometric accuracy without and with additional
GAIA photometry.  The parameters for this event are as in Figure 1,
but the impact parameter $u$ is $2.5$.}
\label{fig:astrophoto}
\end{figure}

\subsection{Photometry and Astrometry}

Figure~\ref{fig:astrophoto} shows the uncertainty in the relative
parallax as a function of astrometric accuracy for events with
astrometry alone and events with astrometry and photometry.  The
latter curves are calculated using the covariance analysis, but now
incorporating the photometric measurements as well
\begin{equation}
b_{ij}=\sum_{k=1}^N \sigmaa^{-2}\
\frac{\partial\theta_\alpha(t_k)}{\partial a_i} \frac{\partial
\theta_\alpha(t_k)}{\partial a_j}
+ \sigma_{\rm A}^{-2}\
\frac{\partial A(t_k)}{\partial a_i} \frac{\partial
A(t_k)}{\partial a_j},
\end{equation}
where $A$ is the photometric amplification and
\begin{equation}
\sigma_{\rm A} = 10^{0.4\sigmap} -1.
\end{equation}
The figure makes the point that just by using GAIA's photometry, it is
possible to improve the accuracy of the parameter estimation by a
factor of $\sim 10$. This however will only be viable for the $\sim
3\%$ of events for which GAIA records a number of datapoints on the
photometric lightcurve (see Section 5.1).

Can GAIA trigger detection of microlensing events which can
subsequently be followed-up on with ground-based telescopes? This is a
real challenge. GAIA provides a one-dimensional stream of astrometric
data, and so must operate for roughly two years to gain enough
accuracy on parallactic motions.  After two years, preliminary
astrometric fits may be constructed, but their accuracy will be
limited. The best opportunities for detecting events will come in the
latter part of the mission, when the iterative astrometric solution
becomes more accurate.  Hence, there is a need for at least one
dedicated 1m class telescope (or better still a world-wide network of
such telescopes) to monitor fields in the bulge and spiral arms so
that the datasets can be correlated with GAIA's database.  The error
is reduced by typically better than a factor of 10 when astrometric
data is supplemented by densely-sampled lightcurves from the ground
(see e.g., Gould \& Salim 1999).

\subsection{The FAME mission}

A related satellite is the {\it Full-Sky Astrometric Mapping Explorer}
or FAME~\footnote{http://www.usno.navy.mil/FAME/}, which is scheduled
for launch in 2004 and has a mission lifetime of 2.5 years. It is
based on the same principles as the {\it Hipparcos} satellite. It is
less ambitious than GAIA, as it monitors mainly sources within about
2.5 kpc of the Sun.  FAME will provide astrometry on all objects with
$6 \lta V\lta 16$.  For a source with $V =10$, the accuracy of
individual measurements $\sigmaa$ is $750 \muas$ and the mission
target accuracy is $36 \muas$. By comparing with
Table~\ref{table:astroacc}, we see that, roughly speaking, the
accuracy of FAME's individual measurements is a factor of 10 worse
than that of GAIA.

The all-sky averaged astrometric optical seen by FAME is $\sim 2
\times 10^{-6}$. There are $\sim 10^8$ stars with $6< V <16$, which
means that FAME will measure $\sim 200$ astrometric microlensing
events (with the characteristics that they have a $5\sqrt{2} \sigmaa$
deviation and that the maximum occurs during the mission lifetime).
This number is low, because the mission lifetime is short and the
individual measurements are noisy.  It is already clear that the high
quality sample will be too small to deduce any useful information
about the local mass function.

\section{Conclusions}

GAIA is an ambitious astrometric survey satellite that is planned for
launch no later than 2012 by the European Space Agency (see ESA 2000;
Perryman et al. 2001). GAIA scans the whole sky, performing
multi-colour and multi-epoch photometry and spectroscopy, as well as
astrometry.  The reference frame is tied to a global astrometric
reference frame defined by extragalactic objects.  For bright sources,
GAIA has the capabilities to perform individual astrometric
measurements with microarcsecond accuracy.  Even for objects as faint
as $V \approx 20$, the positions, proper motions and parallaxes will
be recovered to within at worst a few hundred microarcsecs at the end
of the mission.

GAIA can measure microlensing by the small excursions of the light
centroid that occur during microlensing. We use a stringent definition
of an astrometric microlensing event as one with a significant
variation ($5 \sqrt{2} \sigmaa$) of the centroid shift, together with
a closest approach during the lifetime of the GAIA mission. The
all-sky averaged astrometric microlensing optical depth is $\sim 2.5
\times 10^{-5}$. This means that $\sim 25000$ sources will exhibit
astrometric microlensing events with such characteristics during the
course of the mission.  GAIA can also measure microlensing by the
photometric brightening that accompanies low impact parameter events.
The all-sky averaged photometric optical depth is $\sim 10^{-7}$, so
there are $\sim 3600$ photometric microlensing events during the five
year mission lifetime, most of which are undetectable because of the
poor sampling.  Consequently, very few astrometric events are measured
photometrically as well. In fact, only $\sim 1260$ astrometric events
have at least one datapoint sampled on the photometric
lightcurve. Only for $\sim 427$ of the astrometric events will GAIA
itself obtain enough photometric datapoints to improve substantially
the characterisation of the event. Let us again emphasise that these
numbers only refer to numbers of microlensing events with specified
characteristics. Any statement as to the actual number of microlensing
events that GAIA will find must take into account the number of false
detections thrown up by the identification algorithm.

Our event statistics refer only to the signal provided by stellar
lenses in the Galactic disk. We have not taken into account either
stellar lenses in the Galactic bulge or dark objects in the halo
(popularly called Machos). This is reasonable, as the
cross-section for astrometric microlensing favours nearby lenses.  The
most valuable events are those for which the Einstein crossing time
$\tE$, the angular Einstein radius $\thetaE$ and the relative parallax
of the source with respect to the lens $\pi_{\rm sl}$ can all be
inferred from GAIA's datastream. The mass of the lens then follows
directly. If the source distance is known -- for example, if GAIA
itself measures the source parallax -- then a complete solution of the
microlensing parameters is available. Of these quantities, it is the
relative parallax that is the hardest to obtain accurately. A
covariance analysis is used to follow the propagation of errors and
establish the conditions for recovery of the relative parallax. This
happens if the angular Einstein radius $\thetaE$ is large and the
Einstein radius projected onto the observer plane $\rE \sim 1$ au so
that the Earth's motion about the Sun gives a substantial
distortion. It is also aided if the source is bright so that GAIA's
astrometric accuracy is high and if the duration of the astrometric
event is long so that GAIA has time to sample it fully. These
conditions favour still further lensing populations that are very
close.

Monte Carlo simulations are used to establish the characteristics of
the $\sim 10 \%$ of events for which GAIA can recover the mass of the
lens to good accuracy. The typical lens distance is $\sim 50$ pc and
the typical source distance is $\sim 300$ pc. The highest quality
events seen by GAIA are overwhelmingly dominated by very local lenses.

We conclude that one of the major scientific contributions of
microlensing studies with GAIA will be the determination of the mass
function in the solar neighbourhood. Of course, direct mass
measurements are presently possible just for binary stars with
well-determined orbits. Microlensing is the only technique which can
measure the masses of individual stars. GAIA is the first instrument
with the ability to survey the astrometric microlensing signal
provided by nearby lenses. We have used Monte Carlo simulations to
show that GAIA can reconstruct the mass function in the solar
neighbourhood from the sample of its highest quality events. This works
particularly well for masses exceeding $\sim 0.3 \msun$. Below $0.3
\msun$, the reconstructed mass function tends to underestimate the
numbers of objects, as the highest quality events are biased towards
larger angular Einstein radii.

If there are local populations of low mass black holes, or very cool
halo and disk white dwarfs or very old brown dwarfs, then they will
have easily eluded detection with available technology.  However, the
astrometric microlensing signal seen by GAIA will be sensitive to
local populations of even the dimmest of these stars and the darkest
of these objects.  GAIA is the first instrument that has the potential
to map out and survey our darkest neighbours.

\section*{Acknowledgments}
We wish to thank Andy Gould, Shude Mao and HongSheng Zhao for a number
of insightful comments, as well as Michael Perryman and Tim de Zeeuw
for encouragement and information about the GAIA satellite. We are
particularly indebted to Erik H{\o}g, who provided detailed, helpful
and constructive criticisms of early versions of this paper.

{}

\label{lastpage}
\end{document}